\documentclass[twocolumn,showpacs,preprintnumbers,amsmath,amssymb,superscriptaddress,10pt,aps,prb]{revtex4-2}
\usepackage{epsfig,amsopn}
\usepackage{graphicx}
\usepackage{epstopdf}
\usepackage{braket}
\usepackage{amsmath,amssymb}
\usepackage{amsthm}
\usepackage{enumerate}
\usepackage{xcolor}
\usepackage[colorlinks=true,linktoc=page,linkcolor=magenta,citecolor=magenta]{hyperref}
\newcommand*{\balancecolsandclearpage}{%
	\close@column@grid
	\cleardoublepage
	\twocolumngrid
}

\newcommand\la{\langle}
\newcommand\ra{\rangle}

\newcommand\bs{\boldsymbol}

\newcommand\s{\sigma}

\newcommand{\RNum}[1]{\uppercase\expandafter{\romannumeral #1\relax}}
\usepackage{colortbl}
\begin{document}
	\author{Rekha Kumari}
	\email{rekha@iitk.ac.in}
	\affiliation{Department of Physics, Indian Institute of Technology Kanpur, Kanpur 208016, India}
	\author{Dibya Kanti Mukherjee}
	\affiliation{Department of Physics, Indiana University, Bloomington, Indiana 47405, USA}
	\author{Arijit Kundu}
	\email{kundua@iitk.ac.in}
	\affiliation{Department of Physics, Indian Institute of Technology Kanpur, Kanpur 208016, India}
	\title{Fermi-arcs mediated transport in surface Josephson junctions of Weyl semimetal}
	
	\begin{abstract}
		This study presents Fermi-arcs mediated transport in a Weyl semimetal thin slab, interfacing two $s$-wave superconductors. We present detailed study with both time-reversal and inversion symmetry broken Weyl semimetals under grounding, orbital magnetic fields, and Zeeman fields. An orbital magnetic field induces energy level oscillations, while a Zeeman field give rise to the periodic anomalous oscillations in the Josephson current. These anomalous oscillations correlate with the separation of Weyl nodes in momentum space, junction length, and system symmetries. Additionally, we present an explanation by scattering theory modeling the Fermi-arcs as a network model.
	\end{abstract} 
	\maketitle
	
	\section*{Introduction}
	
	Weyl semimetals (WSMs) are three-dimensional topological systems with accidental bulk degeneracies (Weyl points), near which electrons follow Weyl equations~\cite{ lv2015experimental,lv2015observation,yang2015weyl,PhysRevX.5.011029,xu2016observation,tanaka2012experimental,lv2017observation}.These Weyl points act as monopoles of the Berry curvature in momentum space with a fixed `chirality' (a quantum characteristic determined by the Berry flux enclosed by a closed surface surrounding the node). These Weyl points always occur in pairs due to a no-go theorem~\cite{nielsen1981absence1,nielsen1981absence2}. The realization of a WSM-phase requires either the breaking of time-reversal symmetry (TRS) or inversion symmetry (IR). When TRS and IR coexist, a pair of degenerate Weyl points may emerge~\cite{hasan2011three,burkov2016topological,jia2016weyl}. 
	
	Surface states of Weyl semimetals, often called \textit{Fermi arc} states, arise due to the topological nature of the underlying bands~\cite{huang2015weyl,xu2015discovery1,xu2015discovery2}. These Fermi arcs connect the pairs of Weyl nodes of opposite \textit{chirality}. The requirement that Fermi arcs connect Weyl Fermions of opposite chirality, also leads to them being (pseudo-)spin polarized~\cite{qi2010quantum,zahid2015topological}. In a time-reversal broken (inversion broken) minimal Weyl system, an odd (even) number of such Fermi arcs would be present in the surface Brillouin zone. Being quasi-1D in nature, Fermi arcs are naturally excellent candidates for correlation physics~\cite{moll2016transport,wang2017quantum,resta2018high,jia2016weyl,zheng2021andreev,uchida2014andreev}, although the presence of bulk states cannot be disregarded in describing their properties.
	 
	The chiral nature of Weyl nodes gives rise to unique bulk transport properties in Weyl semimetals~\cite{Vafek_2014,Witczak_Krempa,Hosur_2012}. The separation of Weyl points with opposite chirality enables charge transfer between them when subjected to parallel electric and magnetic fields, a phenomenon known as chiral anomaly. In WSMs, the charge density at an individual Weyl point is not conserved; the application of parallel $E$ and $B$ fields induces the movement of charges from one Weyl point to its counterpart with opposite chirality~\cite{Adler_1969,Ashby_2014}. This charge pumping effect induces a chemical potential difference between the paired Weyl points, termed as chiral charge imbalance or chirality imbalance. Furthermore, at the interface of a superconductor (SC), the transport properties of Weyl electrons have been also investigated. It has been argued that, although in some cases the Andreev reflection process can be blocked~\cite{fu2019josephson}, considering all allowed cases reveals the underlying nature of Weyl systems through Andreev reflection in an SC-Weyl-SC junction, oscillations of the Josephson current carry signatures of momentum-separated Weyl nodes~\cite{lee2014local,li20184pi,li2019zeeman,uddin2019chiral,khanna2016chiral,khanna20170,li2020fermi}. 

	In the case of a slab-geometry Weyl semimetal, when the Fermi energy is near the neutrality point, the Fermi surface consists of these surface states (with a higher density on one surface or another), as well as small bulk Fermi surfaces~\cite{zhang2016quantum,qi2010quantum,zahid2015topological}. These surface states are characterized by a non-vanishing Chern number when projected onto a two-dimensional Brillouin zone. This zone is perpendicular to the momentum that connects pairs of Weyl points. The Josephson transport properties of these helical surface states may give rise to unique transport signatures. 
	
	This work focuses on surface Josephson transport in Weyl semimetal slabs, considering two distinct electronic configurations: one with time-reversal symmetry breaking, characterized by 2-node and one Fermi arc features, and the other with inversion symmetry breaking, characterized by 4-node and two Fermi arcs~\cite{mccormick2017minimal,chen2016superconducting}. In high surface-to-bulk ratio WSM geometries, Fermi-arc electrons serve as the primary Josephson current carriers, while bulk states, residing at higher energy levels, have negligible transport contributions. We investigate the Josephson current-phase relationship (CPR) using tight-binding simulations and a network model. To explore the distinct signatures associated with Fermi-arc-mediated transport, we have considered the grounded and not-grounded cases. Furthermore, we study the impact of orbital and Zeeman magnetic fields on the surface transport, which significantly affect Fermi arc length and Weyl node positions within the semimetal slabs.
	
	\begin{figure*}[th!]
		\centering
		\includegraphics[width=.85\linewidth]{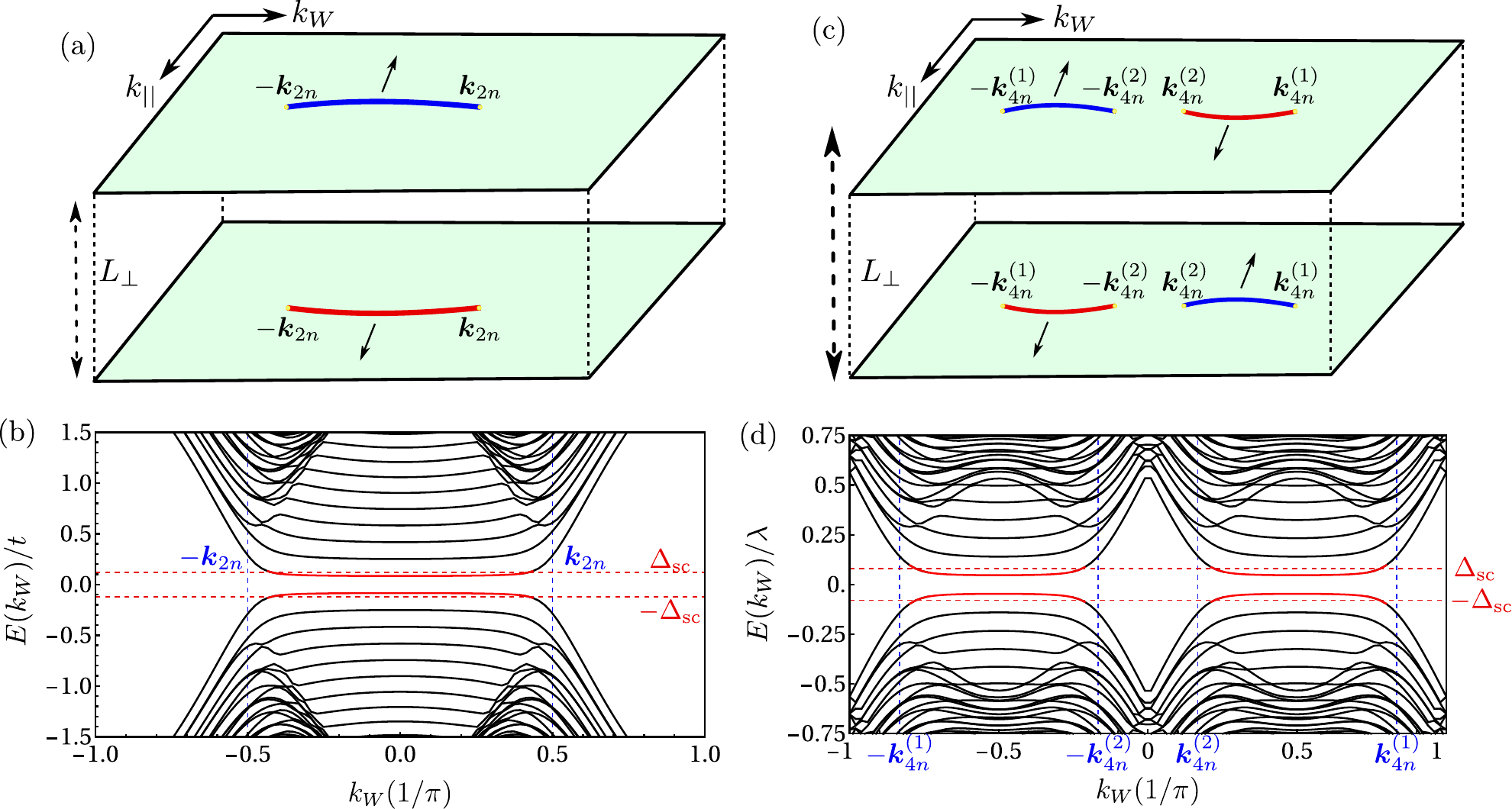}
		\caption{Top: Schematic representations of chiral Fermi arcs in momentum space for the (a) 2-node and (c) 4-node WSMs described by Eq.~(\ref{eq:Ham2n}) and Eq.~(\ref{eq:Ham4n}), respectively. Bottom: The band structure of WSM slab as a function of lattice momentum $k_W$, the Weyl nodes (shown by vertical gridlines) in the bulk are connected by edge states, for both the (b) 2-node and (d) 4-node cases. The parameters are set as follows : $N_{\parallel}=18, N_{\perp}=18$, $m_{2n}=2.0$, $t_x=1, t=1,$ $k_0=\pi/2$, $m_{4n}=0.5$, and $\lambda=1$. The red-marked edge states denote confined states within the superconducting gap ($\pm\Delta_{\rm sc}$), as indicated by horizontal grid-lines. This corresponds to the case of later Josephson transport, where WSM slabs are connected to superconducting reservoirs to form the Josephson junction. 
		}
		\label{fig:E_orb1}
	\end{figure*}
	This paper is structured as follows: In Sec.~\ref{sec2}, we present the electronic band structure and characteristics of Fermi arcs in both 2-node and 4-node Weyl Semimetals. This section also provides details of system setups, along with the effects of applied orbital and Zeeman magnetic fields on the positioning of Weyl nodes and the length of Fermi arcs. In Sec.~\ref{sec:Results}, we analyze numerical and analytical results of the surface Josephson currents under various scenarios mentioned in Sec.~\ref{sec2}. Subsequently, in Section~\ref{sec:conclusions}, we have presented the significance of our work, along with its broader implications, and conclusions drawn from this study.  In the Appendix, Sec.~\ref{sec:Numdetails} and Sec.~\ref{sec:Symmetry-Analysis} outline the tight-binding Hamiltonian for the system along with the computational details of current using the non-equilibrium Green's function formalism and the symmetries of the system, respectively.
	\section{System setup}\label{sec2} 
	\subsection*{Electronic configurations}
	We study two lattice models of WSMs that involve breaking of time reversal and inversion symmetries. Minimal models of them are characterized by the presence of one (two) Fermi arcs and two (four) weyl-nodes in the bulk spectrum, respectively. 
	
	The first one is a 2-band model~\cite{mccormick2017minimal} that describes electrons in a simple cubic lattice, given by the Hamiltonian
	\begin{eqnarray}\label{eq:Ham2n}\nonumber
		H_{2n} & = &m_{2n}(2 - \cos k_{\parallel} - \cos k_{\perp}) \s_x  + 2t_x(\cos k_W - \cos k_0)\\ 
		& &  \s_x + 2t\sin k_{\parallel}\s_y + 2t\sin k_{\perp}\s_z.
	\end{eqnarray}
	Here, $\s_i$ represents Pauli spin matrices, and the Weyl nodes are located at $\pm \bs{k}_{2n}=(k_{\parallel}, k_W, k_{\perp}) = (0, \pm k_0, 0)$ with $t_x=t$, $m_{2n}=2t$. All momentum here are made unitless by scaling with inverse lattice-spacing ($a_0$). The time-reversal symmetry (given by  $\mathcal{T}_{2n} = i\s_y\mathcal{K} |_{\vec{k} \rightarrow -\vec{k}}$, $\mathcal{K}$ being complex conjugation) is broken, while inversion symmetry ($\mathcal{P}_{2n} = \s_x$) is preserved. Throughout the text, we set $\hbar=1$.
	
	\begin{figure*}[th!]
		\centering
		\includegraphics[width=.75\linewidth]{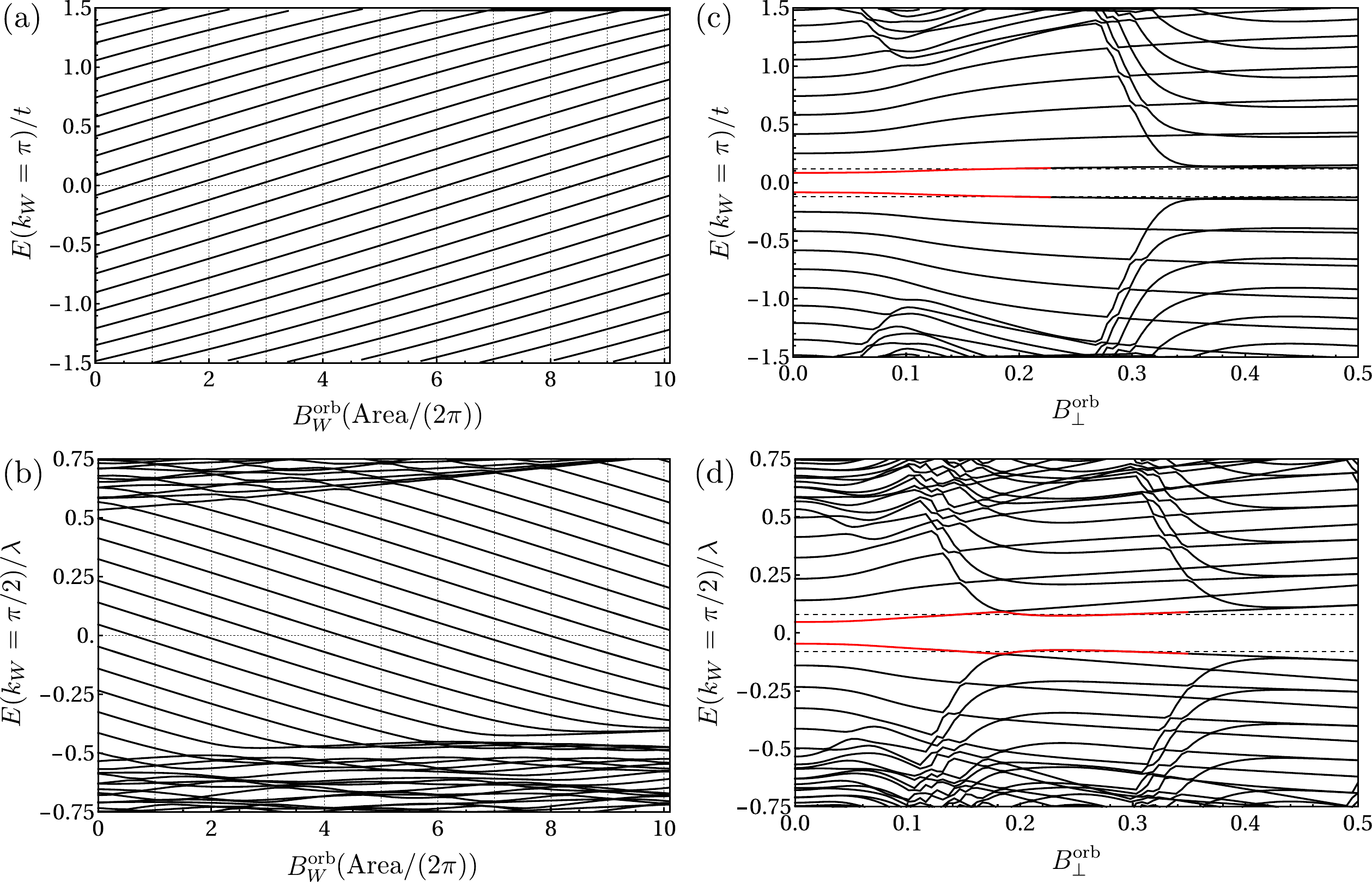}
		\caption{Top: The energy spectrum dependence is shown for the orbital field along \(x_{W}\) (in units of `$\rm Area/2\pi$') 
		in (a) and (b) for the 2-node and 4-node cases, respectively. `$\rm Area$' is the area of the WSM slab given by $(L_{\parallel}-1)(L_{\perp}-1)a^2_0$. The energy spectrum dependence is shown for the orbital field along \(x_{\perp}\) in (c) and (d) for the 2-node and 4-node cases, respectively. Parameters are the same as in Fig.~\ref{fig:E_orb1}.
		}
		\label{fig:E_orb2}
	\end{figure*}
	The second model is a 4-band model~\cite{chen2016superconducting} that describes electrons in a simple cubic lattice with two orbitals per site. The corresponding Hamiltonian is given by:
	\begin{eqnarray}\label{eq:Ham4n}\nonumber
		H_{4n} & = & \lambda s_0(\s_x\sin k_{\parallel}+\s_y \sin k_W+\s_z \sin k_{\perp})+ts_y \s_y \\
		& & (2 + m_{4n} - \cos k_{\parallel} - \cos k_{\perp}).
	\end{eqnarray}
	Here, $s_i$ and $\s_i$ represent the Pauli matrices associated with orbital and spin degrees of freedom, respectively. The parameter $\lambda$ represents the strength of the spin-orbital coupling term and we have set $t$ to $1$. When $m_{4n} > \lambda$ (taking $\lambda>0$), the Hamiltonian corresponds to a trivial insulating phase. For $m_{4n} = \lambda$, the model Hamiltonian exhibits two Dirac nodes located at $(k_{\parallel},k_{W},k_{\perp})=(0,\pm\pi/2,0)$. For $|m_{4n}| < \lambda$, the Hamiltonian corresponds to a Weyl semimetal phase. In this case, each Dirac node splits into two Weyl nodes, situated at $\pm \bs{k}_{4n}^{(1)}=\{ 0,\pi-\sin^{-1}(|m_{4n}/\lambda|),0\}, \pm \bs{k}_{4n}^{(2)}= \{0,\pm\sin^{-1}(|m_{4n}/\lambda|),0\}$. In this 4-node case, time-reversal symmetry is preserved, while inversion symmetry is broken. The inversion and time-reversal symmetry operators are denoted by $\mathcal{P}_{4n} = s_x\s_x$ and $\mathcal{T}_{4n} = is_0\s_y \mathcal{K}|_{\vec{k}\rightarrow-\vec{k}}$, respectively.
	
	We consider our system (described either by Eq.~\ref{eq:Ham2n} or Eq.~\ref{eq:Ham4n}) with finite dimensions along \(x_{\parallel}\) (longitudinal) and \(x_{\perp}\) (transverse) directions. The dimensions of this WSM slab are \(L_{\parallel}=(N_{\parallel}-1)a_0\) and \(L_{\perp}=(N_{\perp}-1)a_0\) for length and width, where \(N_{\parallel}\) and \(N_{\perp}\) represent lattice sites along their respective directions, and \(a_0\) is the lattice spacing. For the rest of the text, we set $a_0=1$, such that our lengths are scaled by the lattice spacing.
	
	
	For the 2-node WSM, a single Fermi arc is present on the surface Brillouin zone, as shown schematically in the Fig.~\ref{fig:E_orb1}(a). The corresponding energy spectrum of the WSM slab is depicted in Fig.~\ref{fig:E_orb1}(b) where the presence of these helical surface states are visible. In the case of the 4-node WSM, two chiral Fermi arcs are present, as shown in Fig.~\ref{fig:E_orb1}(c). This leads to the emergence of two types of helical surface states and the corresponding energy spectrum of the WSM slab is illustrated in Fig.~\ref{fig:E_orb1}(d).
	
	\begin{figure}[b]
		\centering
		\includegraphics[width=.85\linewidth]{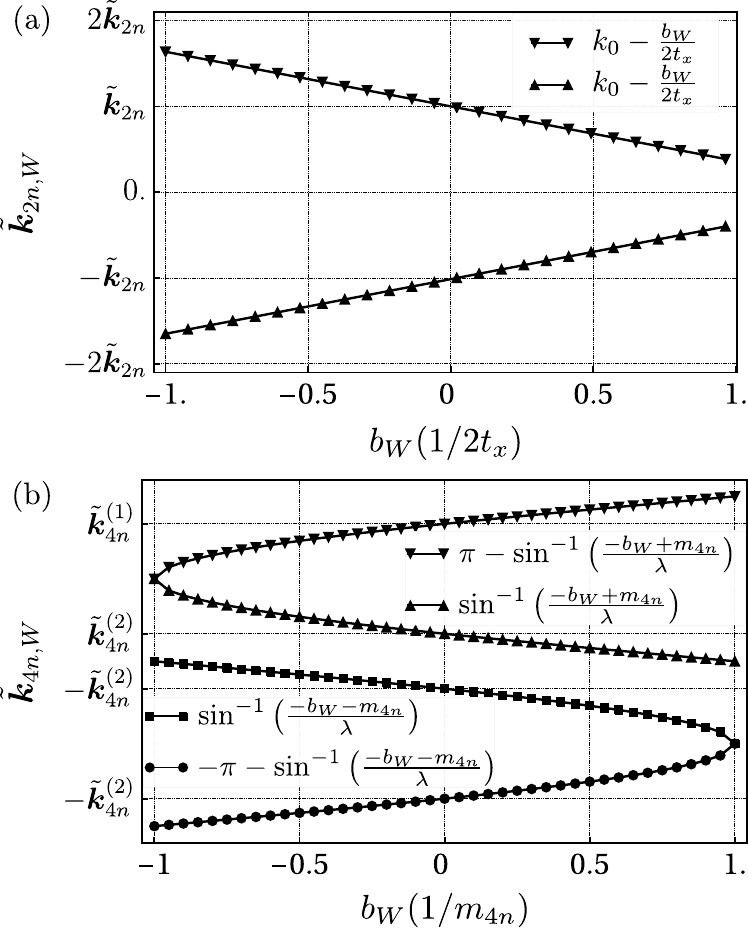}
		\caption{Shift in the positions of Weyl node along the $k_W$ axis with respect to $b_W$ Zeeman field for (a) 2-node and (b) 4-node WSMs. 
		}
		\label{fig:E_ZF1}
	\end{figure}
	\textit{Effect of Orbital Magnetic Fields:} An orbital magnetic field is introduced by $\vec{k} \rightarrow \vec{k} - e\vec{A}^{\text{orb}}$, where $\vec{A}^{\text{orb}}$ represents the vector potential associated with the orbital magnetic field. In a slab geometry, when an orbital magnetic field is applied in the $x_{W}$ direction and is given by the vector potential $\vec{A}^{\rm orb}_{W} = B^{\rm orb}_W x_{\perp}\hat{x}_{\parallel}$, a repeating structure in energies appears, as shown in Fig.~\ref{fig:E_orb2}(a) and Fig.~\ref{fig:E_orb2}(b) for the 2-node and 4-node WSMs, respectively. The oscillation period (of $B^{\rm orb}_W$) of this recurring energy structure is given by $\frac{2\pi}{\text{Area}}$, here, `Area' represents the surface area of the slab, expressed as \((L_{\parallel}-1)(L_{\perp}-1)a^2_0\) and \(n\) is an integer. For each value of \(n\), there exists a crossing of an energy level with the Fermi level.
	
	\begin{figure*}[th!]
		\centering
		\includegraphics[width=.99\linewidth]{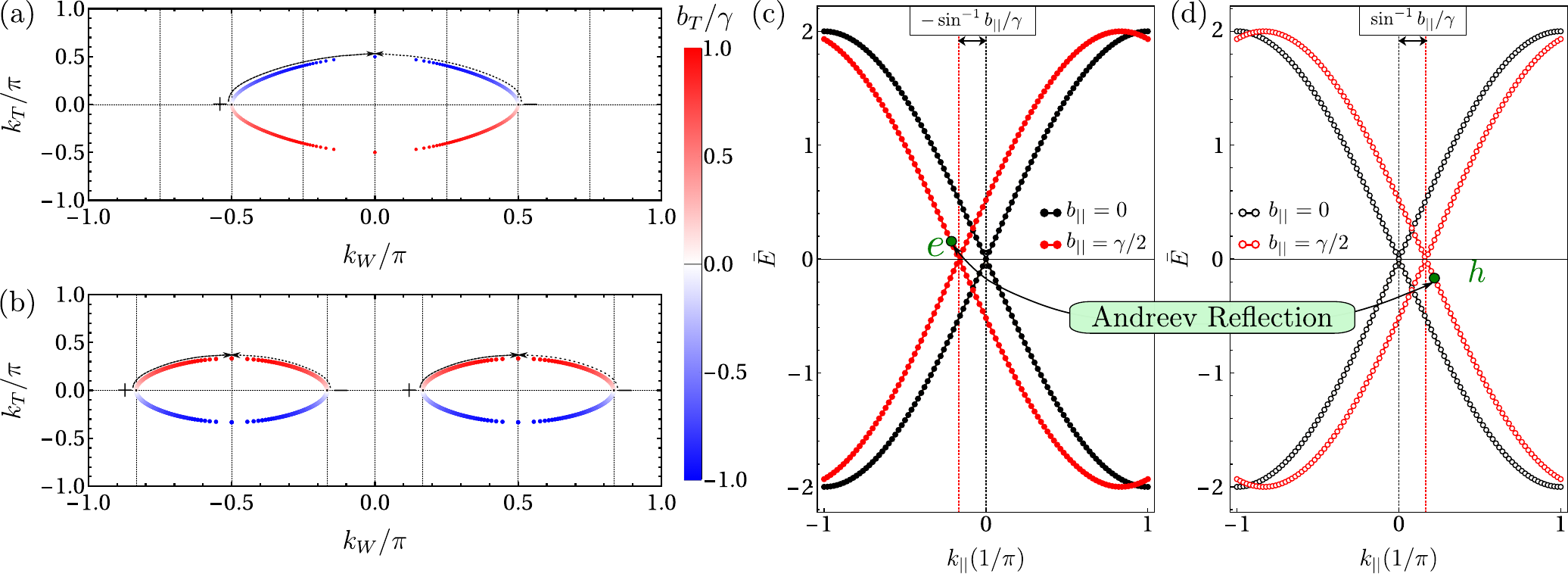}
		\caption{Zeeman field-induced displacement of Weyl nodes in 2-node and 4-node WSMs (indicated by arrows). Shifted positions of Weyl nodes in the $(k_W, k_T)$ plane  in presence applied Zeeman field $B_T$, shown for the (a) 2-node and (b) 4-node WSMs, respectively. The colorbar indicates the strength of the Zeeman field. Here $T$ is either parallel ($\parallel$) or perpendicular ($\perp$) directions. The Zeeman field $b_{\parallel}$ induces opposing shifts in (c) electronic and (c-d) hole bands. The Andreev-reflection processes, shown in (c-d), lead to formation of Andreev bound states from the Fermi arcs. There can be two possible kinds, One Fermi Arc Andreev (OFAR) processes, when these are within the states of the same Fermi arc, and Two Fermi Arc Andreev (tFAR) processes, when these are within the states of two different Fermi arcs. In the case of 2-node WSMs, $\gamma=2t_x, \bar{E}=E/t$, with $k_W$ given by $\bs{\tilde{k}}_{2n,W}$ in (c) and $-\bs{\tilde{k}}_{2n,W}$ in (d). For 4-node WSMs, $\gamma=\sqrt(1-m_{4n}^2), \bar{E}=E/\lambda$, with $k_W$ given by $\bs{\tilde{k}}^{(1)}_{4n,W}$ in (c) and $-\bs{\tilde{k}}^{(1)}_{4n,W}$ in (d).
		}
		\label{fig:E_ZF2}
	\end{figure*}
	Furthermore, when an orbital magnetic field is applied in the $x_{\perp}$ direction and is given by the vector potential $\vec{A}^{\rm orb}_{\perp} = B^{\rm orb}_{\perp} x_{\parallel}\hat{x}_{W}$, it results in a shift of the energy levels of the surface states, as illustrated in Fig.~\ref{fig:E_orb2}(c) and Fig.~\ref{fig:E_orb2}(d) for the 2-node and 4-node Weyl semimetals, respectively.
	
	\textit{Effect of Zeeman Fields:} To account for the influence of a Zeeman field, the Hamiltonians for the 2-node (given in Eq.~(\ref{eq:Ham2n})) and 4-node (given in Eq.~(\ref{eq:Ham4n})) incorporate the following terms, respectively:
	%
	\begin{eqnarray}\label{eq:Hzf2n}
		H_B^{2n} &=& b_W\s_x+b_{\parallel}\s_y+b_{\perp}\s_z\\
		H_B^{4n} &=& b_{\parallel}\s_x+b_{W}\s_y+b_{\perp}\s_z.
	\end{eqnarray}
	%
	In presence of these additional Zeeman field terms in the Hamiltonian the modified position of the Weyl nodes in the case of 2-node and 4-node WSM are given by: 
	\begin{align}\label{eq:Weylnodes_zf_2n}
		& \bs{\tilde{k}}_{2n} = \left(\sin^{-1}\alpha_{\parallel}, \pm f_{\alpha_{\parallel},\alpha_W,\alpha_{\perp}}, \sin^{-1}\alpha_{\perp}\right);\\\nonumber\label{eq:Weylnodes_zf_4n}
		& \bs{\tilde{k}}_{4n} = \left(\sin^{-1}\beta_{\parallel}, \left\{\sin^{-1}\left(\pm g_{\beta_{\parallel},\beta_W,\beta_{\perp}}\right), \right.\right. \\
		& \hspace{14mm} \left.\left.\pm\pi-\sin^{-1}\left(\pm g_{\beta_{\parallel},\beta_W,\beta_{\perp}}\right)\right\}, \sin^{-1}\beta_{\perp}\right);
	\end{align}
	respectively. Here, in the above equations:
	\begin{align}\nonumber
		&f_{\alpha_{\parallel},\alpha_{\perp}} = k_0-\alpha_W-\bar{m}_{2n}\big(2-(1-\alpha_{\parallel}^2)^{1/2}-(1-\alpha_{\perp}^2)^{1/2}\big),\\\nonumber 
		&g_{\beta_{\parallel},\beta_{\perp}} = \bar{m}_{4n}-\beta_W+2-(1-\beta_{\parallel}^2)^{1/2}-(1-\beta_{\perp}^2)^{1/2}.
	\end{align}
	$\alpha_{||(\perp)} = b_{||(\perp)}/2t$,  
	$\beta_{||(\perp)} = b_{||(\perp)}/\lambda$, $\alpha_W=b_W/2t_x$, $\beta_W=b_W/\lambda$, $\bar{m}_{2n}=m_{2n}/2t_x$, and $\bar{m}_{4n}=m_{4n}/\lambda$ .
	
	\begin{figure}[ht!]
		\centering
		\includegraphics[width=.75\linewidth]{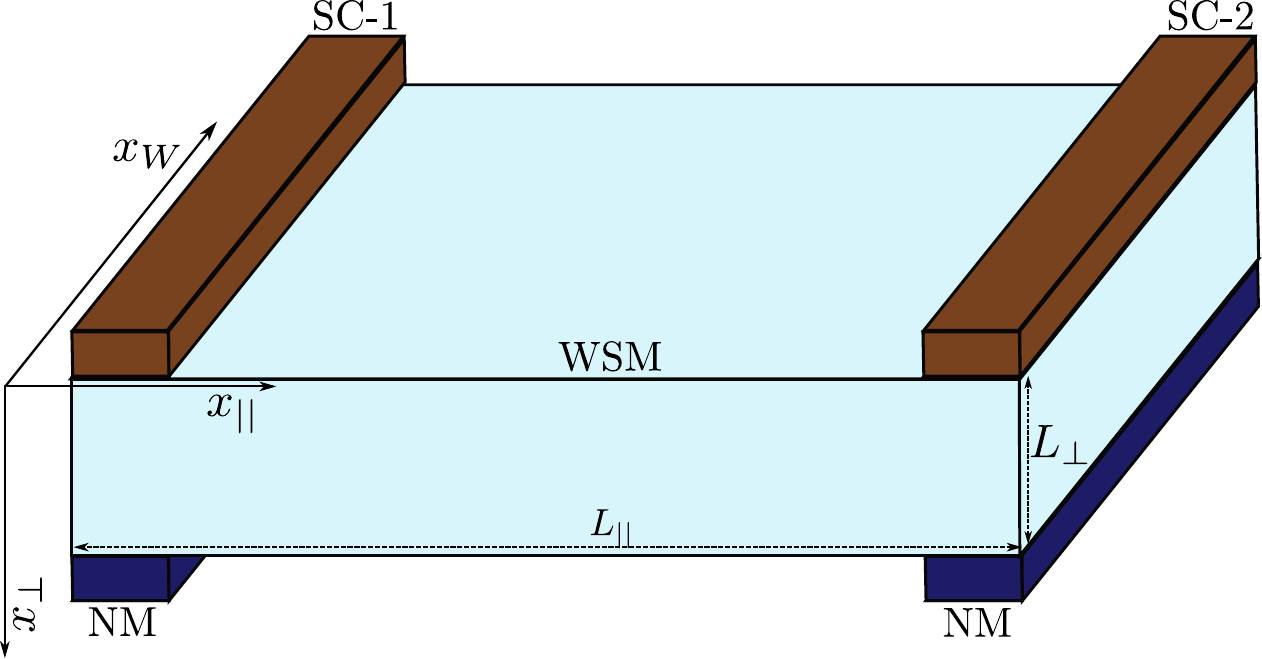}
		\caption{System setup: s-wave superconducting leads, spaced $L_{\parallel}$ apart, attach to WSM's top surface. The bottom surface connected to normal leads and the top and bottom surfaces spaces $L_{\perp}$ apart.}
		\label{fig:setup}
	\end{figure}
	In Fig.~\ref{fig:E_ZF1}(a) and Fig.~\ref{fig:E_ZF1}(b), the variation in the length of Fermi arc(s) is presented as a function of \(b_W\) for the 2-node and 4-node cases, respectively. In the 2-node case, it is observed that the Zeeman field changes the length of the Fermi arc. In contrast, within the 4-node case, the Zeeman field term induces relative difference in the lengths of Fermi arcs.

	\begin{figure*}[ht!]
		\centering
		\includegraphics[width=0.75\linewidth]{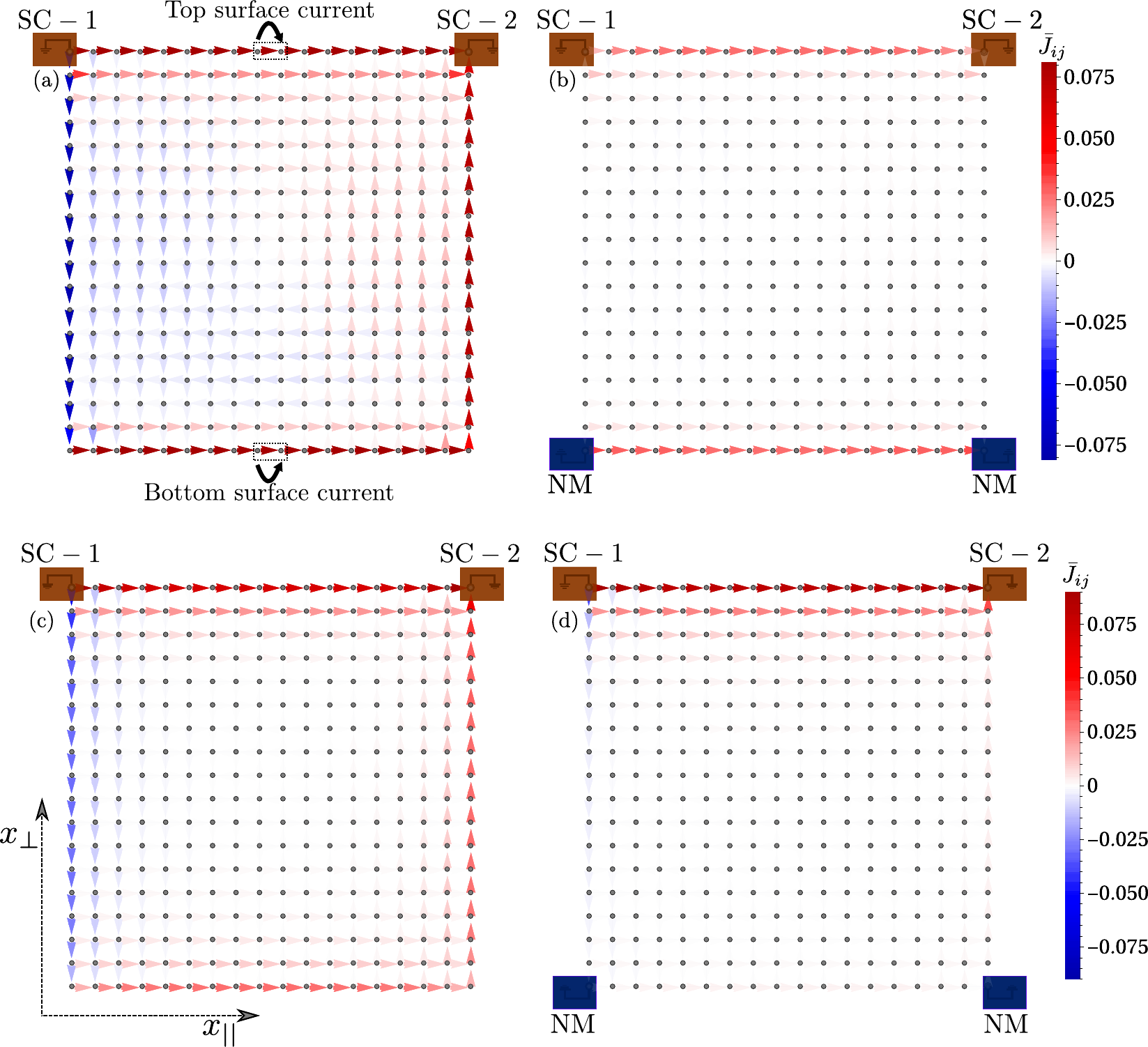}
		\caption{The bond current distribution on the square lattice of the 2-node WSM is depicted. 
			Panels (a) and (b) depict the difference \(J_{ij}(\delta\phi=\pi/2) - J_{ij}(\delta\phi=0)\) in the 2-node case. Panels (c) and (d) depict the \(J_{ij}(\delta\phi=\pi/2)\) in the 4-node case. Panels (a), (c), and (b),(d) correspond to the grounded and non-grounded scenarios, respectively. Colors define the magnitude of the Josephson current, visualized through the color bar. The parameters used are: $t_{sc}=1$, $\mu_{\text{sc}}=0$, $\Delta_{\text{sc}}^{2n}=0.125$, $\Delta_{\text{sc}}^{4n}=0.1$, $t_{\lambda}=1$ and other parameters are same as given in Fig.~\ref{fig:E_orb1}.
		}
		\label{fig:4node}
	\end{figure*}
	
	In Fig.~\ref{fig:E_ZF2}(a) and Fig.~\ref{fig:E_ZF2}(b), we illustrate the displacement of the Weyl node in the \((k_W,k_{T})\) plane as a function of the Zeeman field \(b_{T}\) for the 2-node and 4-node cases, respectively. The colorbar denotes the magnitude of the Zeeman field. We observe that as the Zeeman field \(b_{T}\) increases, the separation of Weyl nodes along the \(k_W\) direction decreases, consequently leading to a reduction in the length of the Fermi arc. Specifically, for \(b_{T}=\gamma\), we observe the annihilation of Weyl nodes with opposite chirality, as depicted by arrows in Fig.~\ref{fig:E_ZF2}(a-b).  The reduction in the length of the Fermi arcs results in a decrease in the number of surface states participating in Josephson transport. Additionally, \(b_{T}\) induces shifts in the locations of Weyl nodes along the \(k_{T}\) direction.
	
	Fig.~\ref{fig:E_ZF2}(c-d) presents the shift in energy dispersion near Weyl nodes as a function of lattice momentum \(k_{\parallel}\), for different values of the Zeeman field \(b_{\parallel}\). In the 2-node WSM, Fig.~\ref{fig:E_ZF2}(c) and Fig.~\ref{fig:E_ZF2}(d) illustrate the shift in the electronic and hole bands near Weyl nodes \(\bs{k}_{2n,W}\) and \(-\bs{k}_{2n,W}\), as only a single Fermi arc is present.
	
	\noindent In the 4-node case, (c) and (d) show the shift in the electronic and hole bands near Weyl nodes of the first Fermi arc \(\bs{k}^{(1)}_{4n,W}\) and \(-\bs{k}^{(1)}_{4n,W}\), in addition to this the hole bands near the Weyl node \(-\bs{k}^{(2)}_{4n,W}\) from the second Fermi arc accumulate the same shift as shown in (d). 
	
	When the WSM slabs are connected with the superconducting reservoirs, these low energy states contributes in the transport through Andreev reflections. In the case of 2-node WSM, in the presence of superconducting reservoirs, a top-edge electron with spin-up (spin-down) undergoes Andreev reflection as a hole state with spin-down (spin-up) along the bottom edge. This process, termed the ``One Fermi Arc Andreev Reflection (OFAR) Process,'' involves the transfer of a charge of \(2e\) from left to right reservoirs. However, when the system is grounded, the introduction of normal reservoirs induces decoherence, significantly reducing the total current.
	
	In the case of 4-node WSM, in addition to the `OFAR-Process,' another Andreev reflection process, termed the `Two Fermi Arc Andreev Reflection (TFAR) Process,' occurs due to the presence of the second Fermi arc. In the TFAR process, a spin-up (spin-down) electron on the top edge reflects as a spin-down (spin-up) hole on the same edge, incorporating two Fermi arcs and one surface. This process involves both Fermi arcs and the top edge.
	
	\begin{figure*}[hbt!]
		\includegraphics[width=0.9\textwidth]{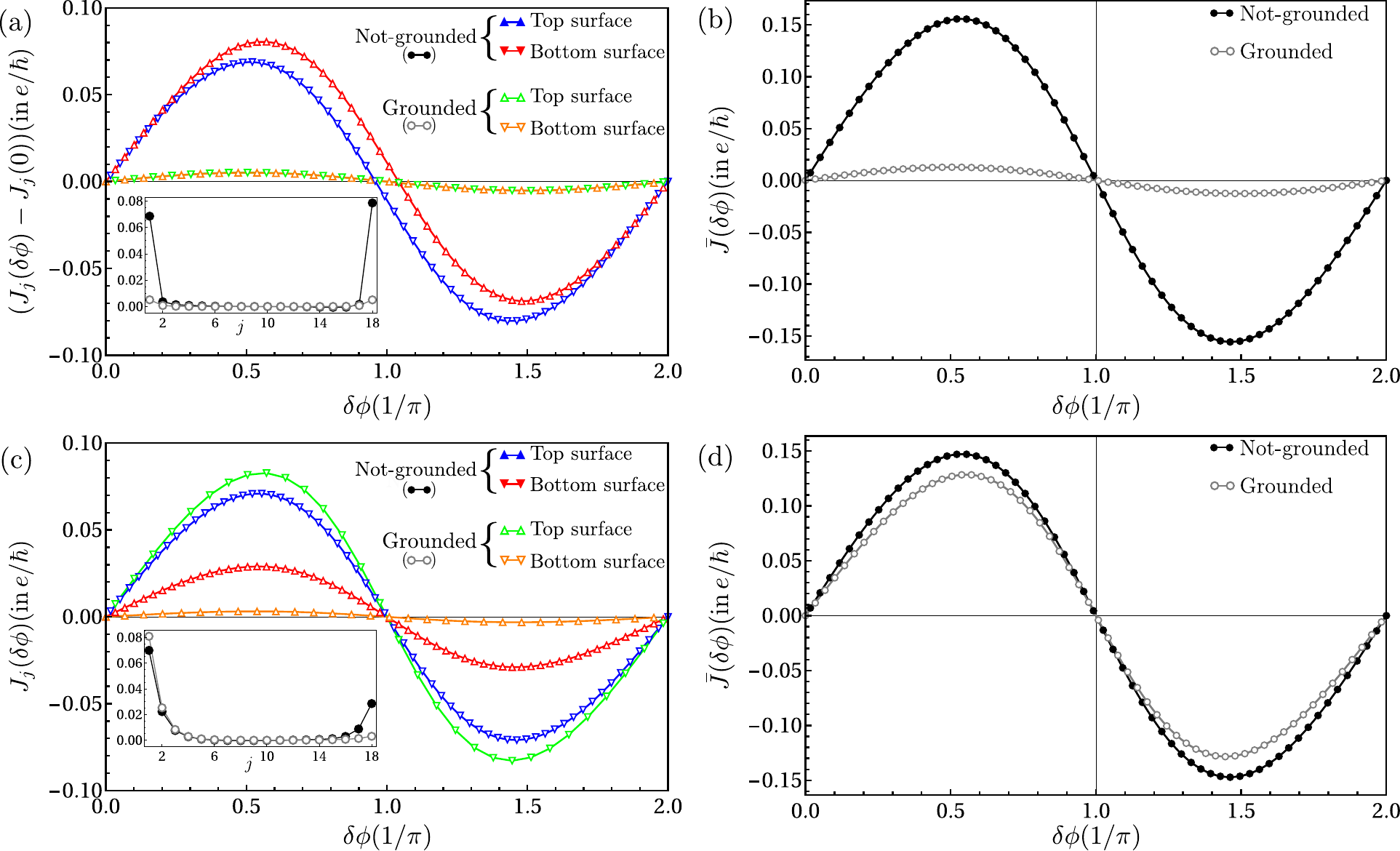}
		\caption{Current-Phase Relations (CPR) :For the 2-node case, (a) and (b) show surface and summed Josephson currents as functions of superconducting phase difference \(\delta\phi\) for both not grounded and grounded cases. Similarly, in the 4-node case, (c) and (d) show surface and summed Josephson currents as functions of superconducting phase difference \(\delta\phi\) for not grounded and grounded scenarios. Insets provide the current contributions from different layers along $x_{\perp}$ direction for each case. These currents are computed using Eq.~(\ref{eq:j2n_||}), Eq.~(\ref{eq:j2n_perp}), Eq.~(\ref{eq:j4n_||}) and Eq.~(\ref{eq:j4n_perp}).
		}\label{fig:CPR}
	\end{figure*}
	
	\subsubsection*{Setup details}
	To study surface transport in Weyl semimetals, we have considered two geometrical setups. The first case is the `not-grounded' configuration, where at the top surface (\(j=N_{\perp}\)) of the WSM slab two superconducting reservoirs are connected, while the bottom surface is not connected to any reservoirs. 
	
	The second case is the `grounded' configuration, where, in addition to the two superconducting reservoirs connected at the top surface, the bottom surface (\(j=1\)) is connected to normal reservoirs, as illustrated in Fig.~\ref{fig:setup}. 
	
	To explore transport properties of edge states, we adjust the superconducting gaps to selectively confine the minimum-energy surface states, as shown in Fig.~\ref{fig:E_orb1} (highlighted in red). Subsequent sections elaborate on the specifics of the tight-binding Hamiltonians governing the superconducting and normal reservoirs, as well as the Weyl semimetal slabs. Moreover, the methodology encompasses a comprehensive description of the bond and net Josephson current computations employing non-equilibrium Green's function (NEGF) techniques.
	
	\section{Numerical Results:}\label{sec:Results}	
	\subsection*{Surface Josephson effect:}
	The bond current distribution in the case of 2-node and 4-node WSM are shown in Fig.~\ref{fig:4node}. We have computed these bond current using the Non-Equilibrium Green's Function (NEGF) formalism mentioned in Sec.~\ref{sec:Numdetails}. 
	
	For the 2-node case, as time-reversal symmetry is already broken, a persistent surface current is there in the absence of a superconducting phase difference. 
	The bond currents between the sites $\{(i,j)\rightarrow(i+1,j)\}$ and $\{(i,j)\rightarrow(i,j+1)\}$ can be computed as 
	\begin{align}
		J_{(i,j)\rightarrow(i+1,j)}&=-\frac{2e}{\hbar}\text{Im}\Big[\sum_{\eta\eta'}t_{ij\eta,i+1 j\eta'} \left\langle c^{\dagger}_{i,j,\eta}c_{i+1,j,\eta'}\right\rangle\Big],\\
		J_{(i,j)\rightarrow(i,j+1)}&=-\frac{2e}{\hbar}\text{Im}\Big[\sum_{\eta\eta'}t_{ij\eta,i j+1\eta'} \left\langle c^{\dagger}_{i,j,\eta}c_{i+1,j,\eta'}\right\rangle\Big],
	\end{align}
	respectively. Here, $t_{ij\eta,i+1 j\eta'}$ incorporates the details of the hopping elements between sites $(i,j)$ and $(i+1, j)$ and $\eta,\eta'$ incorporates the onsite degree of freedom indices. $\langle\ldots\rangle$ is the thermal average taken over the reservoir's states. Using this description we have computed the bond current between for sites $\{(N_{\parallel}/2,N_{\perp})\rightarrow (N_{\parallel}/2+1,N_{\perp})\}$ and $\{(N_{\parallel}/2,1)\rightarrow (N_{\parallel}/2+1,1)\}$ are termed as top and bottom surface(layer) currents, respectively, as shown by arrows in Fig.~\ref{fig:4node}(a). 
	
	Additionally, Fig.~\ref{fig:4node}(a) and Fig.~\ref{fig:4node}(b) show the distribution of bond currents on this 2D square lattice for the phase \(\delta\phi=\pi/2\), obtained by subtracting the persistent currents for the not-grounded and grounded cases, respectively. For a fixed \(k_W\in(-k_0,k_0)\), each edge exclusively accommodates a distinct type of helical surface state. We observe that the prominent current flow occurs along the top (bottom) edges of the system, and there is a suppression in bond currents for the grounded scenario as a result of dephasing induced by normal reservoirs. The color of the colorbar represents the amplitude of the bond currents.
	
	For the 4-node case, Fig.~\ref{fig:4node}(c) and \ref{fig:4node}(d) illustrate the bond current distribution on the square lattice in the non-grounded and grounded cases, respectively, at a phase difference of $\delta\phi=\pi/2$. In contrast to the two-node case, there are enhancements in the top-layer current and a reduction in the bottom-layer current, underscoring the significance of grounding in this case. 
	
	Fig.~\ref{fig:CPR} depicts the variation of top and bottom surface Josephson currents, along with the net Josephson current, as functions of the superconducting phase difference. 
	
	Fig.~\ref{fig:CPR}(a) exhibits the surface currents (after subtracting this persistent flow from \(J_{j}(\delta\phi)\),) and Fig.~\ref{fig:CPR}(b) illustrates the net Josephson current as a function of the superconducting phase difference \(\delta\phi=\phi^L-\phi^R\), respectively. In both the not-grounded and grounded cases, there exists a net current flow along the \(\hat{x}_{\parallel}\) direction. Notably, the predominant contributions to the net Josephson current originate from the top and bottom surfaces. However, in the grounded case, we observe a significant reduction in the magnitudes of the net and layer currents compared to the not-grounded case.
	
	Fig.~\ref{fig:CPR}(c) and \ref{fig:CPR}(d) depict the surface and net Josephson currents for the 4-node Weyl semimetal as functions of the superconducting phase difference. In the non-grounded scenario, analogous to the 2-node case, the principal contribution to the net current arises from the top and bottom surfaces. In the grounded configuration, the top surface current intensifies while the bottom surface current diminishes, resulting in a total current of comparable magnitude to the non-grounded case. This is in contrast to the 2-node case, where grounding suppresses the net current.
	
	In the 2-node case, the Josephson current is facilitated through the occurrence of OFAR-processes, involving both surfaces. The introduction of the grounding lead induces decoherence in the system, leading to the suppression of the net current. Conversely, in the 4-node Weyl semimetal, in addition to OFAR processes, TFAR-processes occur due to the presence of the second Fermi arc. Given that TFAR processes involve both Fermi arcs and only the top surface, grounding affects OFAR processes involving the bottom surface but favors TFAR processes. This results in an increase in the top-layer current, a decrease in the bottom-layer current, and the maintenance of the net current magnitude for the 4-node Weyl semimetal case.
	
	\begin{figure*}[th!]
		\includegraphics[width=0.85\textwidth]{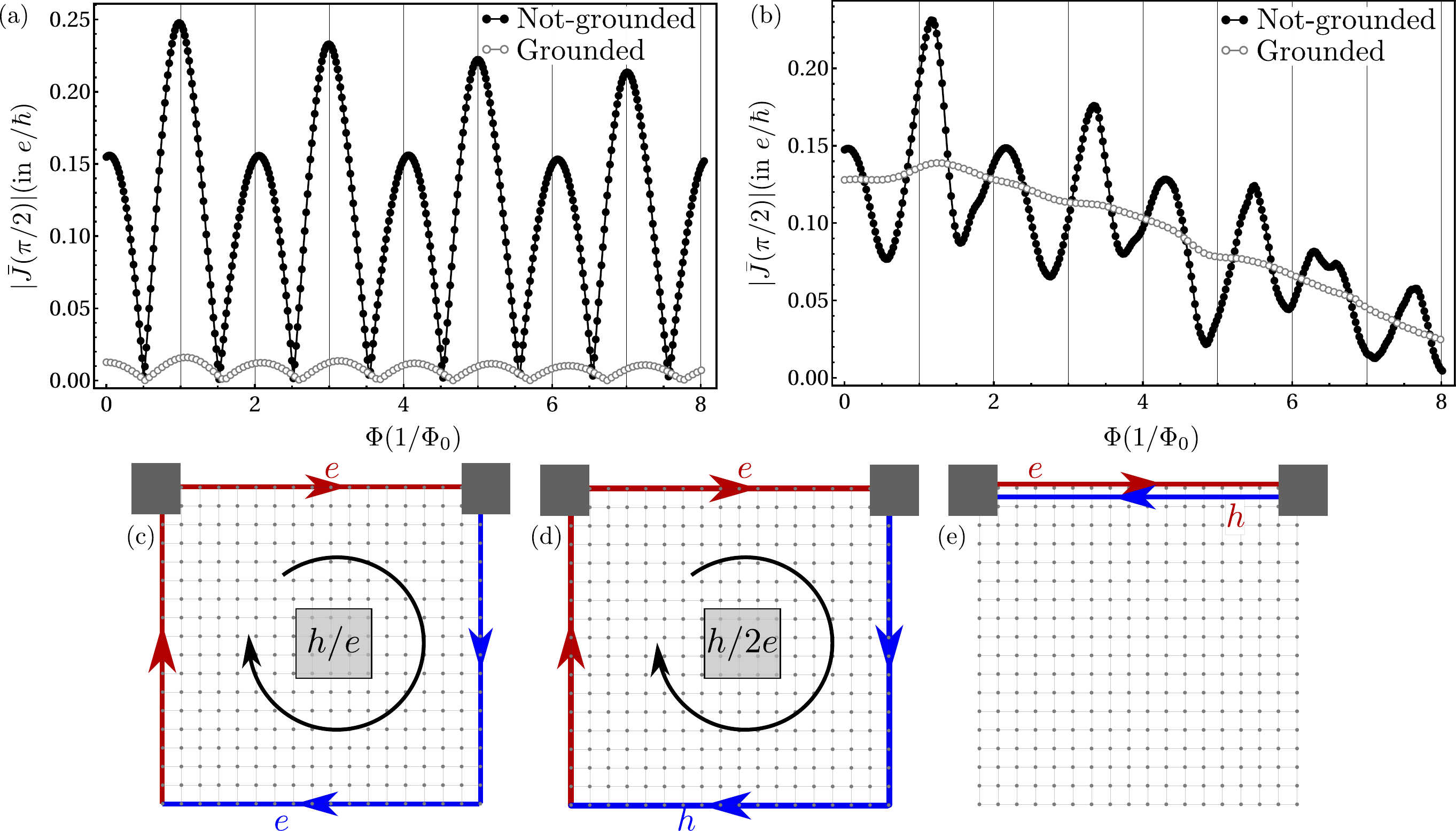}
		\caption{Oscillations in the summed Josephson current (at a phase difference of $\pi/2$) as a function of the orbital magnetic field strength $B_W$ for both the (a) 2-node and (b) 4-node WSM cases in a non-grounded and grounded setups. Parameters are same as Fig.~\ref{fig:4node} Conducting chiral paths for charge transfer: (c) The circulating persistent current flow of charge \(e\). (d) OFAR process and (e) TFAR process for a \(2e\) charge from the left to right reservoirs.}
		\label{fig:Jorb1}
	\end{figure*}
	\begin{figure}[th!]
		\includegraphics[width=0.42\textwidth]{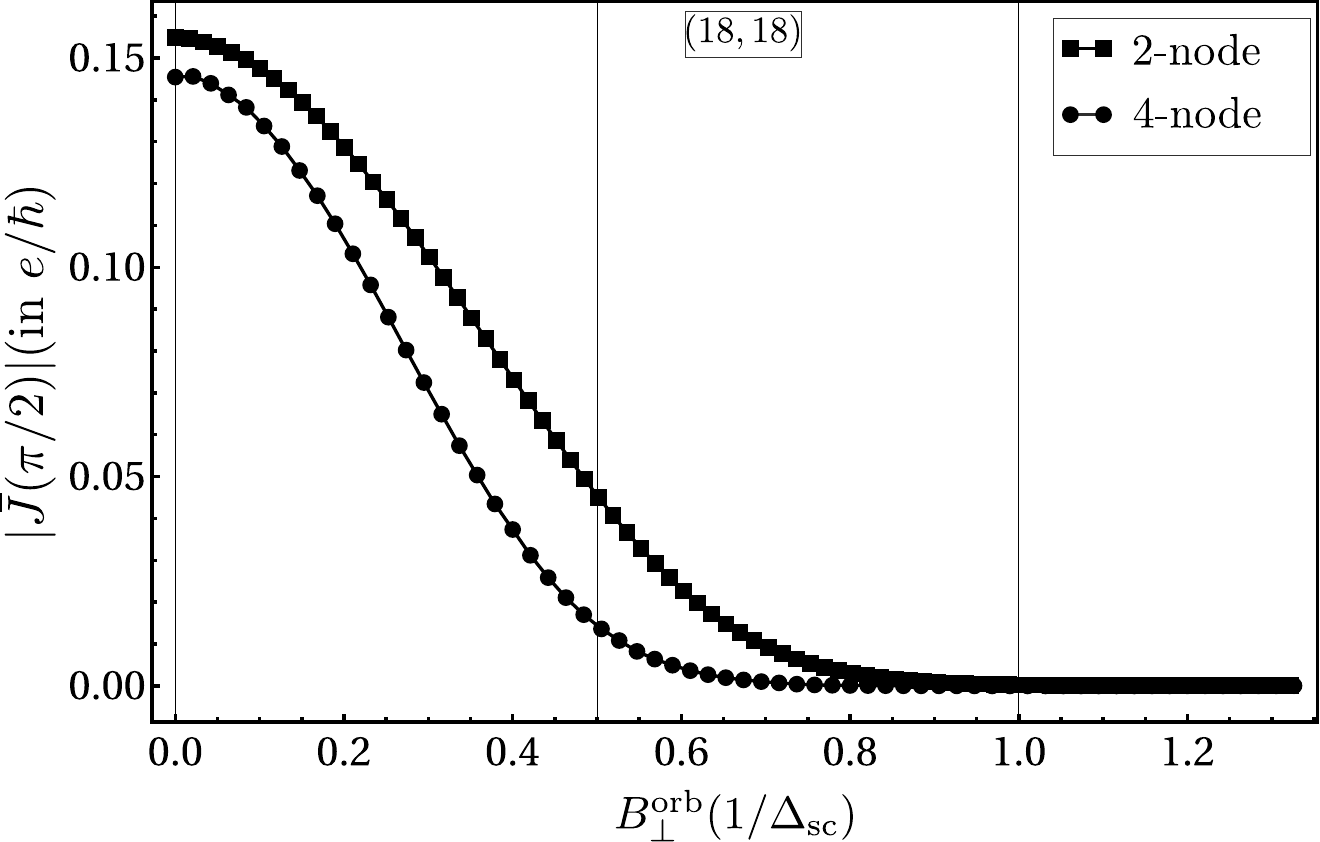}
		\caption{Variation in the summed Josephson current as a function of applied orbital magnetic field $B^{\text{orb}}_{\perp}$ for both the 2-node and 4-node WSM cases in the non-grounded case.}
		\label{fig:Jorb1a}
	\end{figure}
	\begin{figure*}[ht!]
		\includegraphics[width=0.95\textwidth]{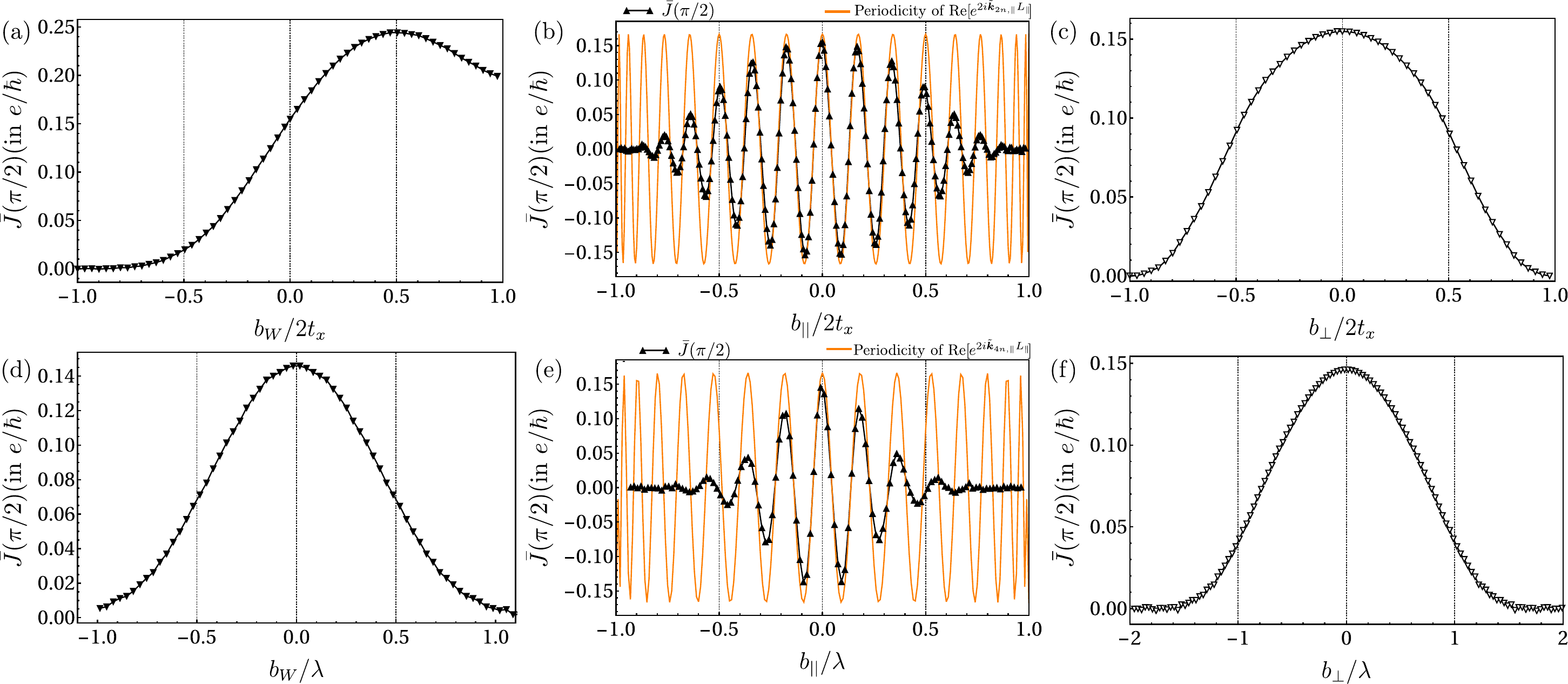}
		\caption{Variation in summed Josephson current (at a phase difference of $\pi/2$) in the (top row) 2-node and (bottom row) 4-node WSM cases with respect to applied Zeeman fields along the $b_{W}$, $b_{\parallel}$, and $b_{\perp}$ directions, in panels (a), (d); (b), (e); and (c), (f). The parameters used in these plots are identical to those in Fig.~\ref{fig:E_orb1} and Fig.~\ref{fig:CPR}. Parameters $M_{2n}$ and $M_{4n}$ are $2t_x$ and $\sqrt{1-m_{4n}^2}$, respectively.
		}
		\label{fig:JZF1}
	\end{figure*}
	In the insets of Fig.~\ref{fig:CPR}(b) and Fig.~\ref{fig:CPR}(d), contributions from different \(j\)-layers in grounded and not-grounded cases for 2-node and 4-node WSMs are presented, highlighting minimal bulk current contributions.
	
	\subsection*{Effect of orbital magnetic fields} 	
	Fig.~\ref{fig:Jorb1}(a) and Fig.~\ref{fig:Jorb1}(b) present the dependence of the net Josephson current (for a fixed phase difference \(\delta\phi=\pi/2\)) on the orbital magnetic field \(B^{\text{orb}}_{W}\) for the 2-node and 4-node Weyl semimetals, respectively. In the 2-node case, as depicted in Fig.~\ref{fig:Jorb1}(a), the net Josephson current exhibits an oscillatory dependence on \(\Phi\) (in units of \(1/\Phi_0\)), where \(\Phi = B^{\text{orb.}}_W \times \text{Area}\), \(\text{Area} = L_{\parallel} L_{\perp}a^{2}_0\) is the area of the slab, and \(\Phi_0 = \frac{h}{2e}\) is the superconducting flux quantum (in units of \(\hbar=1\) and \(e=1\)).
	
	We observe that whenever \(\Phi=n\Phi_0\), where \(n\) is an integer, there is a peak in the net Josephson current. Additionally, we observe a substantial difference in the peak heights of the net Josephson current at even and odd values of \(n\). Also, we find that grounding the bottom surface in the 2-node case suppresses the amplitude of the net Josephson current while preserving the oscillatory behavior.
	
	Similar oscillations have also been observed in the 4-node case, as illustrated in Fig.~\ref{fig:Jorb1}(b), but with a less clear oscillation period. Additionally, in contrast to the 2-node case, grounding the bottom surfaces in this case results in the complete disappearance of these oscillations.
	
	Our numerical results align with similar findings in theoretical works~\cite{baxevanis2015even,tkachov2015quantum,meier2016edge,sun2023crossover} and experimental works~\cite{pribiag2015edge,de2018h,wang2020evidence} and can be elucidated as follows. In the context of a surface Josephson junction, when an applied magnetic field is present along the \(x_{\perp}\)-direction, the phase difference accumulates an additional phase shift along different \(x_{\perp}\)-layers due to the applied magnetic field, given by \(\frac{2\pi B_{W}x_{\perp}L_{\parallel}}{\Phi_0}\), where \(\Phi_0=\frac{h}{2e}\) is the superconductive flux quantum, with \(h\) being the Planck constant, and \(2e\) representing the Cooper pair charge. In the current-phase relation expressions, we observed that the maximum current flow is along the edges of the top (bottom) layers of the Weyl semimetal (WSM). This allows us to express \(J_{c}(x_{\perp})\approx(J^{\text{bot.}}_{c}\delta_{x_{\perp},1}+J^{\text{top}}_{c}\delta_{x_{\perp},N_{\perp}})\), leading to the summed current in the form of: \(\bar{J}_{c} = (J^{\text{top}}_{c}+J^{\text{bot.}}_{c})\cos\left[\frac{\pi \Phi}{\Phi_0}\right]\).
	
	In the 2-node case, the even-odd effect and periodicity of the net Josephson current can be elucidated by considering distinct conductive paths, as illustrated in Fig.~\ref{fig:Jorb1}(c-e). The top and bottom edge helical channels are connected by the left (right) edge channels. The path shown in Fig.~\ref{fig:Jorb1}(c) presents the inherent persistent current flow due to time reversal symmetry breaking, with enclosed magnetic flux \(\frac{h}{e}\). The path shown in Fig.~\ref{fig:Jorb1}(d) presents the charge flow via the OFAR process, with enclosed magnetic flux \(\frac{h}{2e}\). 
	
	These two-dimensional conductive paths exist for each \(k_W\), giving rise to the even-odd Fraunhofer-like patterns in the net Josephson current. In the grounded case, these 2D paths get obstructed by the induced decoherence in the system, and the net Josephson current amplitude gets suppressed while the oscillating nature is maintained, as shown in Fig.~\ref{fig:Jorb1}(a).
	
	In contrast to this, in the 4-node case, an additional one-dimensional conductive path arises, as shown in Fig.~\ref{fig:Jorb1}(e), showcasing the TFAR process for $2e$ charge flow without enclosing any flux. These contributions yield a less clear oscillation period compared to the 2-node case, as this process adds up a constant contribution to the net current without enclosing any flux. Furthermore, in the grounded case, normal reservoirs obstruct the conductive paths (a-b), and only path (c) contributes to the net current, leading to the complete destruction of even-odd oscillations in the 4-node WSM, as shown in Fig.~\ref{fig:Jorb1}(b).
	
	The variation in net Josephson current as a function of the magnetic field \(B^{\text{orb}}_{\perp}\) is shown in Fig.~\ref{fig:Jorb1a}. We observe that the Josephson current decreases with an increase in the magnetic field. This suppression in net current occurs because the magnetic field \(B^{\text{orb}}_{\perp}\) shifts the surface states to higher energy, as shown in Fig.~\ref{fig:E_orb2}(c) and Fig.~\ref{fig:E_orb2}(d), for the 2-node and 4-node WSMs, respectively.
	
	\subsection*{Effect of Zeeman fields}
	In Fig.~\ref{fig:JZF1} (top row), we present the net Josephson current (for a fixed phase difference $\delta\phi=\pi/2$) as a function of applied Zeeman fields in the 2-node case. As shown in Fig.~\ref{fig:JZF1}(a), the Zeeman field $b_W$ serves as a controlling parameter, causing a reduction (negative values of the field) or amplification (positive values of the field) of the current. This effect results from variations in the magnetic field altering the Fermi arc length, controlling the number of edge states, and subsequently influencing the net current magnitude, as shown in Fig.~\ref{fig:E_ZF1}(a).
	
	In Fig.~\ref{fig:JZF1}(b), we observe that the net Josephson current oscillates as a function of the Zeeman field $b_{\parallel}$. Additionally, the current diminishes as the applied magnetic field $|b_{\parallel}|$ increases, approaching zero for $|b_{\parallel}|\geq2t_x$. These oscillations exhibit periodicity represented by $\theta^{2n}=\bar{\bs{k}}_{2n,\parallel}L_{\parallel}\text{mod}(2\pi)$. 
	The origin of the oscillations in net current lies in opposite momentum shifts induced by the Zeeman field in electronic and hole low-energy band dispersion. The electronic band dispersion undergoes a shift of $\bs{\tilde{k}}_{2n,\parallel}$, while the hole band dispersion undergoes a shift of $-\bs{\tilde{k}}_{2n,\parallel}$, as shown in Fig.~\ref{fig:E_ZF2}(c) and Fig.~\ref{fig:E_ZF2}(d), respectively.
	
	In the 2-node case, the net current flow is governed by the OFAR-Andreev reflection process. When an electron traverses the top edge from the left reservoir to the right, it accumulates a phase of $e^{i\bs{\tilde{k}}_{2n,\parallel}L_{\parallel}}$, and a hole reflects from the right reservoir to the left reservoir along the bottom edge, accumulating the same phase. Consequently, a total phase accumulation of $e^{2i\bs{\tilde{k}}_{2n,\parallel}L_{\parallel}}$ occurs in this process. These accumulated phases result in distinctive oscillations in net current variation as a function of Zeeman field $b_{\parallel}$. In Fig.~\ref{fig:JZF1}(b), the $\text{Im}[e^{2i\bs{\tilde{k}}_{2n,\parallel}L_{\parallel}}]$ function is also plotted to showcase the identical oscillation period between the current and the periodicity of this function.
	
	The decrement in net Josephson current, as shown in Fig.~\ref{fig:JZF1}(b) and Fig.~\ref{fig:JZF1}(c), arises when the Zeeman field is oriented along the $b_{\parallel}$ and $b_{\perp}$ directions, respectively. These fields reduce the length of Fermi arcs, and the number of conductive edge channels decreases, proportional to the Zeeman field strength as shown in Fig.~\ref{fig:E_ZF2}(a). This results in a gradual decrease and eventual zeroing of the current when the Zeeman field strength equals $2t_x$.
	
	In Fig.~\ref{fig:JZF1} (bottom row), we present the net Josephson current as functions of applied Zeeman fields for the 4-node case. In contrast to the 2-node case, as shown in Fig.~\ref{fig:JZF1}(d), the net Josephson current vanishes as a function of the Zeeman field $|b_W|$. This reduction in the current occurs because, in the 4-node case, the relative difference in the length of Fermi arcs increases in the presence of the applied Zeeman field $b_W$. This leads to the suppression of the TFAR process, resulting in a decrement in the net Josephson current amplitude.
	
	In Fig.~\ref{fig:JZF1}(e), we observe that similar to the 2-node case, the net Josephson current exhibits oscillations as a function of Zeeman field $b_{\parallel}$ in this case as well. Analogous to the OFAR processes, in the 4-node case, the TFAR processes also accumulate a total phase shift of $e^{2i\bar{k}_{4n,\parallel}L_{\parallel}}$, as shown in Fig.~\ref{fig:E_ZF2}(c,e). These phase shifts result in oscillations in the variation of the net current as a function of the Zeeman field. The periodicity of these oscillations is given by $\theta^{4n}=2\bar{\bs{k}}_{4n,\parallel}L_{\parallel}\text{mod}(2\pi)$. 
	In this case as well, the net current diminishes as the applied magnetic field $|b_{\parallel}|$ increases, approaching zero for $|b_{\parallel}|>|\lambda\sqrt{1-m_{4n}^{2}}|$.
	
	The suppression in net Josephson current, as shown in Fig.~\ref{fig:JZF1}(e) and Fig.~\ref{fig:JZF1}(f), arises when the Zeeman field is oriented along the $b_{\parallel}$ and $b_{\perp}$ directions, respectively. Similar to the 2-node case, $b_{\parallel(\perp)}$ fields decrease the Fermi arc lengths as shown in Fig.~\ref{fig:E_ZF2}(b). Consequently, the net current gradually decreases and reaches zero when the Zeeman field strength equals $\lambda\sqrt{1-m_{4n}^{2}}$.
	\subsection*{Anomalous currents} 
	
	\begin{figure*}[ht!]
		\centering
		\includegraphics[width=0.85\linewidth]{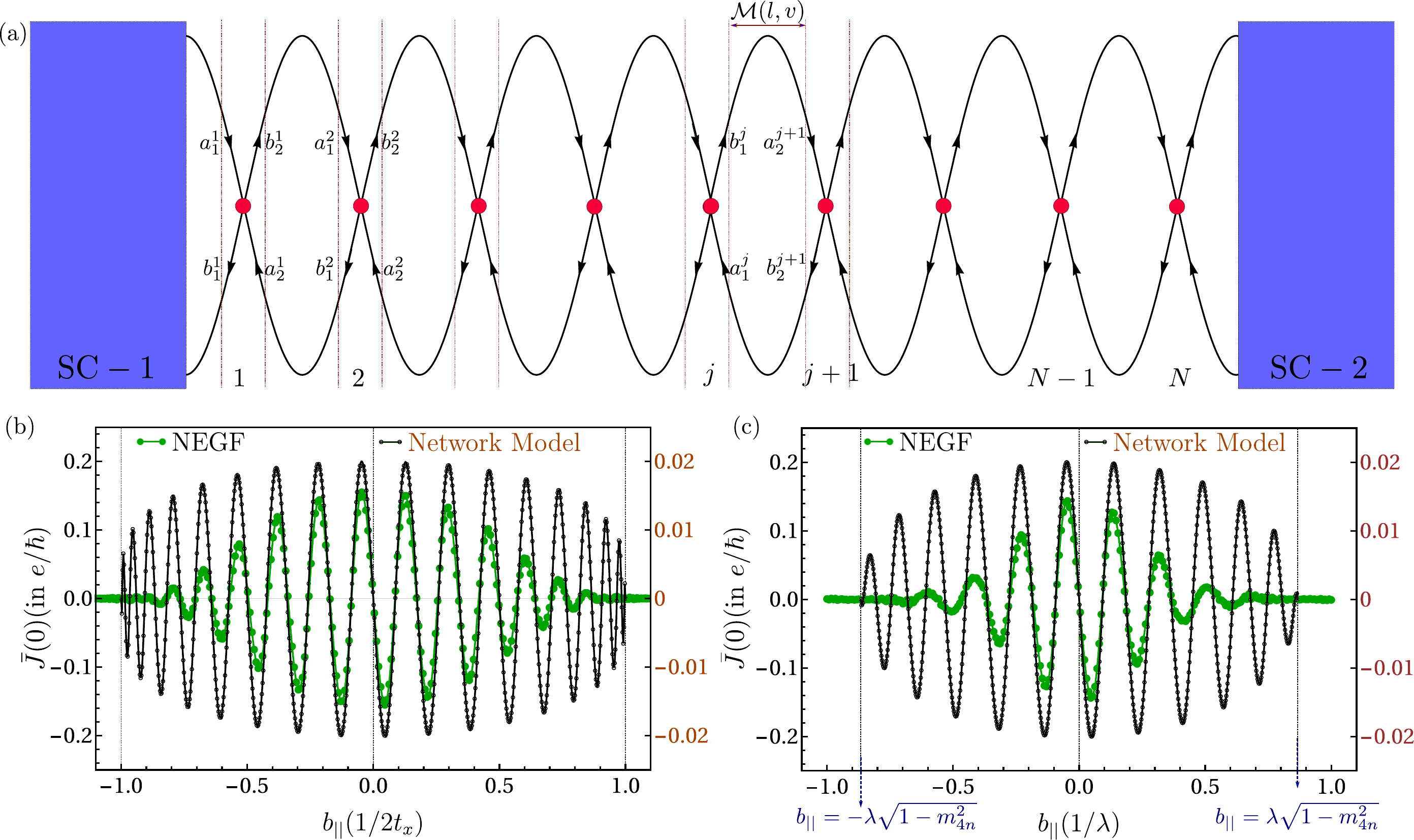}
		\caption{
			Network model to study Josephson junctions: A disordered region coupled with two superconducting regions, labeled as $\text{sc}1$ and $\text{sc}2$. Variation in anomalous Josephson current (at $\delta\phi=0$) is shown for the (a) 2-node and (b) 4-node WSM cases. The graphs display the variation in current with Zeeman fields along $b_{\parallel}$, reflecting $\bs{\tilde{k}}_{2n(4n),||}L_{\parallel}$ oscillation compared to the analytically computed Josephson current, computed using a network-model study. The parameters are the same as those in Fig.~\ref{fig:E_orb1} and Fig.~\ref{fig:CPR}.
		}
		\label{fig:NM}
	\end{figure*}
	Additionally, we also observe that the introduction of a Zeeman field $b_{\parallel}$ leads to the emergence of `anomalous' (computed for $\delta\phi=0$) Josephson current, characterized by similar oscillations in both WSM cases. The variation in this anomalous current as a function of $b_{\parallel}$ is shown in Fig.~\ref{fig:NM}(b-c) for the 2-node and 4-node WSMs, computed using NEGF for system sizes $(18,18)$ and $(18,16)$. In Section \ref{sec:Symmetry-Analysis}, a symmetry analysis has been presented for both the 2-node and 4-node cases. The anomalous current observed can be traced back to the inherent symmetries embedded within the system.
	
	From Eq.~(\ref{eq:slist2n}) and Eq.~(\ref{eq:slist4n}) given in sec.~\ref{sec:Symmetry-Analysis}, we observe that unique symmetry operators are identified for the 2-node and 4-node cases, represented as \(\mathcal{P}_{2n} = \mathcal{\tilde{T}}_{2n} \tilde{\sigma}_y \tilde{\mathcal{R}}_{\perp}\) and \(\mathcal{P}_{4n} = \mathcal{\tilde{T}}_{4n}|_{k_W\rightarrow-k_W} \tilde{s}_z \tilde{\sigma}_x \tilde{\mathcal{R}}_{\perp}\), respectively. This symmetry yields \(\mathcal{P}_{\beta} \mathbb{H}^{\beta}_{\text{full}}(\Phi^L,\Phi^R) \mathcal{P}_{\beta}^{-1} = \mathbb{H}^{\beta}_{\text{full}}(-\Phi^L,-\Phi^R)\), where \(\mathbb{H}^{\beta}_{\text{full}}\) denotes the full Hamiltonian of the Josephson junction (i.e. WSM slab connected with the superconducting reservoirs). Consequently, this symmetry implies \(E(\phi)=E(-\phi)\), ensuring \(I(\phi)=-I(-\phi)\) and thereby establishing the absence of anomalous current.
	
	In the presence of the Zeeman field \(b_{\parallel}\), this symmetry is disrupted, resulting in the emergence of anomalous current in both WSM cases, while for all other cases, this symmetry remains intact. The breakdown of this symmetry gives rise to \(E_{n}(\phi^L,\phi^R) \neq E_{n}(-\phi^L,-\phi^R)\), leading to \(I(\phi) \neq -I(-\phi)\) and consequently causing the presence of anomalous current. In order to gain a further understanding of this anomalous current and its oscillatory behavior, we conducted a network model study, as given in the following.
	
	\subsubsection*{Network-model Analysis}\label{sec:Network-model}
	In this section, we explain results of anomalous Josephson response in presence of $b_{\parallel}$ field using a network model. This approach employs a network representation, as discussed in Refs.~\cite{giuliano2013josephson,brouwer1997anomalous,beenakker1991universal} of the Fermi arcs. Fig.~\ref{fig:NM}(a) illustrates the network model for a Josephson junction, where two s-wave superconductors are coupled with a chain of 1D scatterers. These scatterers are labeled \(\{1, 2, \ldots, j, j+1, \ldots, N-1, N\}\) and function as nodes, as shown in the figure.
	
	For this system, the Josephson current can be computed by utilizing the scattering matrix of the system. The scattering matrix of this system is constructed by directly combining the scattering matrices of individual nodes and bonds, denoted as \(\bs s_{\text{node}}\) and \(\bs s_{\text{bond}}\), respectively. The node matrix, denoted as $\bs s_{\text{node}}$, has a block-diagonal structure with scattering matrices $\bs s_i$ of individual nodes having indices $i = 1, 2, \ldots$ along the diagonal. Due to electron-hole decoupling in the normal region, each $\bs s_i$ follows a block-diagonal pattern comprising electron and hole blocks: $\bs s_{i,e}(\omega)$ and $\bs s_{i,h}(\omega) = (\bs s_{i,e}(\omega))^*$. This results in $\bs s_{\text{node}} = \bs s_{1,e}(\omega) \oplus \bs s_{1,h}(\omega) \oplus \bs s_{2,e}(\omega) \oplus \bs s_{2,h}(\omega) \oplus \ldots$. In the 4-node case, these blocks also account for spin and orbital degrees of freedom, given as $\sigma$ and $s$, respectively. In the 2-node case, these blocks only account the spin degree of freedom. The node matrices, denoted as $s_{\text{node}}$, for each node-$j$ has a $2m\times 2m$ (here $m$ denotes the local degree of freedom at each site), electronic scattering matrix $\bs s_j$, which relates incoming and outgoing wave amplitudes of the channels, according to $\bs b^{j+1}=\bs s_j\bs a^{j}$, here $\bs b^{j}=(b^{j}_1,b^{j}_2)^T$ and $\bs a^{j}=(a^{j}_1,a^{j}_2)^T$.
	To compute the Josephson current in this system we have considered a generalized form of scattering system given as:
	\begin{align}
		\bs s_{j}=\begin{pmatrix}
			\cos\theta & \sin\theta\\
			-\sin\theta & \cos\theta\
		\end{pmatrix}\otimes s_0\otimes\tau_0\otimes\s_0
	\end{align}
	The bond matrix, represented as $\bs s_{\text{bond}} = \mathcal{M}(l,v)$, incorporates phase factors for outgoing-to-incoming mode mapping, i.e. $\bs a^{j+1}_{1}=e^{i\epsilon l/\hbar\nu}\mathcal{I}_{m\times m} \bs b^{j}_{2}$ and $\bs a^{j}_{2}=e^{i\epsilon l/\hbar\nu}\mathcal{I}_{m\times m} \bs b^{j+1}_{1}$ in the scattering region. 
	At the interface, the bond matrix contains the Andreev reflections probabilities given as: $\bs a^{\lambda}_{1}=\mathcal{M}_A(\phi^{\lambda})\bs b_{\lambda}$, here $\mathcal{M}_A(\phi^{\lambda})$ is the Andreev reflection matrix; given as $\mathcal{M}_A(\phi^{\lambda})_{s'\eta'\sigma',s\eta\sigma}=i\eta'\alpha(\omega)e^{i\eta'\phi^{\lambda}}\delta_{s',s},\delta_{\tau',\bar{\tau}}\delta_{\s',\bar{\s}}$. Here, $\alpha(\omega)=i(\omega/\Delta_0)+\sqrt{1-(\omega/\Delta)^2}$ and $\lambda=L, R$ represents the left, right superconducting reservoirs, respectively. 
	
	$\phi^{\lambda}$ is the superconducting phase of $\lambda$-t   h reservoir. Indices $(s,s')\in(1,2)$, $(\eta,\eta')\in(+1,-1)$ and $(\s,\s')\in(\uparrow,\downarrow)$ corresponds to orbital, particle-hole and spin degree of indices, respectively. For $|\omega| < \Delta$, the bond matrix $s_{\text{bond}}(\omega)$ is unitary, but for $|\omega| > \Delta$, the Andreev reflection probability $|\alpha(\omega)|^2$ drops below unity due to propagating modes in the superconductor. 
	
	
	As derived in Ref.~\cite{brouwer1997anomalous,beenakker1991universal}, the Josephson current at temperature $\theta$ is then a sum of the logarithmic determinant over Fermionic Matsubara frequencies $\omega_p =(2p + 1)k_{B}\theta$ ($k_{B}$ is the Boltzmann constant):
	\begin{align}\label{eq:J-nm}
		&J_0 = -\frac{k_B\theta}{2e\hbar} \sum_{p=0}^{\infty} \ln \det [1 - s_{\text{node}}(i\omega_p) s_{\text{bond}}(i\omega_p)],\\
		&= \frac{k_B\theta}{2e\hbar} \sum_{p=0}^{\infty} \text{Tr} \left[1 - s_{\text{node}}(i\omega_p) s_{\text{bond}}(i\omega_p)\right]^{-1} s_{\text{node}} \, ds_{\text{bond}}' \bigg|_{i\omega_p}.
	\end{align}
	From the above-mentioned formalism, we have computed the Josephson current in this 1D Josephson junction, by probing the band-dispersion of WSM in presence of applied Zeeman field. 
	In the presence of an applied Zeeman field, the low-energy dispersion of a WSM (energies denoted by $\epsilon$) undergoes a shift based on the strength of the Zeeman field. As illustrated in Fig.~\ref{fig:E_ZF2}, for both the 2-node and 4-node cases, the energy of the low-energy electronic and hole states shift along the momentum axis in opposite direction. This is incorporated as an additional shift in the $\mathcal{M}$ matrices, where electronic and hole parts acquire the same  phases of $\exp(i \sin^{-1}(b_{\parallel}/\gamma)l)$, where $\gamma=2t$ and $\gamma=\lambda$ for 2 and 4-node case, respectively (see Eq.~(\ref{eq:Weylnodes_zf_2n}) and Eq.~(\ref{eq:Weylnodes_zf_4n})). This phase difference gives rise to anomalous current and the oscillation in the Josephson currents.
	
	Additionally, the Zeeman fields displaces the Weyl nodes and reduces the Fermi arc lengths, as depicted in Fig.~\ref{fig:E_ZF2}. Thus, we take this effect into account by multiplying the Josephson current with the length of Fermi-arc for each value of the field $b_{\parallel}$.

	The variation in Josephson current computed using this network model study, as a function of applied Zeeman field $b_{\parallel}$ are also presented in Fig.~\ref{fig:NM}(a) and Fig.~\ref{fig:NM}(b) for an identical junction length $L_{\parallel}=18$ and $k_B\theta=0.1$, in the 2-node and 4-node cases, respectively. Remarkably, our numerical results consistently align with the network model results.\vspace{2mm}

	\section{Conclusion}\label{sec:conclusions}
	In summary, this work emphasizes the importance of surface transport in Weyl Semimetals across various geometrical and electronic configurations. Specifically, we highlight the significance of one and two Fermi-arc reflection processes in surface Josephson current transport. The grounded configuration of WSM serves as a tool to differentiate between one and two Fermi-arc WSMs. Additionally, we demonstrate the impact of orbital magnetic fields on surface Josephson transport, leading to distinct even-odd Fraunhofer oscillations in the net Josephson current based on Fermi arc parity. This property can effectively distinguish between two types of Weyl semimetals in the grounded setups.
	
	Furthermore, we explore the effects of Zeeman fields on transport, acting as parameters to measure and create the gap (by controlling the Fermi arc lengths) and tuning the Weyl node separations. Experimental setups, similar to those in Ref.~\cite{lee2014local,sochnikov2015nonsinusoidal}, can reveal surface transport through various voltage outcomes. Periodic anomalous oscillations of the Josephson current can be probed through Andreev spectroscopy, with the length scales of such variations typically spanning a few tens of nanometers in typical samples. Tuning Weyl node separation in momentum space is achievable by adjusting the Zeeman field~\cite{guo2023zeeman}. These experimental configurations hold promise for generating controlled, periodically manipulable outputs through the application and adjustment of WSM nodes using Zeeman fields.
	\vspace{2mm}
	\section{Acknowledgments} \label{sec:Acknowledgments}
	R. K. acknowledges the use of PARAM Sanganak and HPC 2013, facility at IIT Kanpur. The support and resources provided by PARAM Sanganak under the National Super-computing Mission, Government of India, at the Indian Institute of Technology, Kanpur, are gratefully acknowledged.
	
	\bibliography{Bibliography1.bib}
	\clearpage
	\appendix
	\section{NUMERICAL Details} \label{sec:Numdetails}
	To construct the Josephson junction on the surface of the Weyl semimetal slab, we employ superconducting leads characterized by the Bogoliubov-de Gennes (BdG) Hamiltonian for a one-dimensional \(s\)-wave superconductor in the particle-hole basis. The BdG Hamiltonian (\(\mathbb{H}^{\alpha}_{\rm sc}\)) for a superconducting lead (\(\alpha\)) is written as:
	\begin{eqnarray}\label{eq:Hscnb}\nonumber
		\mathbb{H}^{\alpha}_{\rm sc} &=& \frac{1}{2} \sum_{i=1}^{N_{\rm sc}} \Phi^{\alpha\dagger}_{i} \Big[ \mu_{\rm sc} \tau_z \sigma_0 \Phi^{\alpha}_{i} + \Delta_{\rm sc} \sigma_0 (\tau_x \cos\phi^{\alpha} - \tau_y \sin\phi^{\alpha}) \\
		& &+ t_{sc} \tau_z \sigma_0 \Phi^{\alpha}_{i\pm1} \Big].
	\end{eqnarray}
	Here \(\tau_z\), \(\tau_x\), and \(\tau_y\) are Pauli matrices acting in the particle-hole space. $\alpha=L$ and $R$ represents the left and right superconducting reservoirs, respectively. The Nambu spinor \(\Phi_i^{\alpha\dagger}\) is defined as \((c_{\uparrow}^{\dagger}, c_{\downarrow}^{\dagger}\),\( c_{\uparrow}^{\alpha}, -c_{\downarrow}^{\alpha})_i\), where \(c_{\uparrow,i}^{\dagger,\alpha}\) represents the creation operator for an electronic state at site \(i\) with spin \(\uparrow\) in the \(\alpha\)-th superconductor. The superconducting phase of the \(\alpha\)-superconductor is denoted by \(\phi^{\alpha}\). The terms \(\Delta_{\rm sc}\), \(\mu_{\text{sc}}\), and \(t_{\text{sc}}\) represent the s-wave pairing gap, chemical potential, and nearest-neighbor hopping amplitude in the superconductor, respectively. \(N_{\text{sc}}\) signifies the total number of sites in each superconducting reservoir.
	
	
	To construct the grounded setup, we consider normal reservoirs modeled by the Bogoliubov-de Gennes Hamiltonian for a one-dimensional normal metal in the particle-hole basis, expressed as follows:
	\begin{equation}\label{eq:Hnmnb}
		\mathbb{H}^{\bar{\alpha}}_{\rm nm} = \frac{1}{2} \sum_{i'=1}^{N_{ \rm nm}} \Phi^{\bar{\alpha}\dagger}_{i'} \Big[ \mu_{\rm nm} \tau_z \sigma_0 \Phi^{\bar{\alpha}}_{i'} + t_{\rm nm} \tau_z \sigma_0 \Phi^{\bar{\alpha}}_{i'\pm1} \Big].
	\end{equation}
	Here, \(\bar{\alpha}=L\) and \(R\) represent the left and right normal reservoirs, respectively. \(\Phi^{\bar{\alpha}\dagger}_{i'}=(a^{\dagger}_{\uparrow},a^{\dagger}_{\downarrow}, a_{\uparrow},-a_{\downarrow})^{\bar{\alpha}}_{i'}\) and \( a_{\uparrow,i}^{\dagger\,\bar{\alpha}}\) represents the electronic creation operator at site \(i\) and with spin \(\uparrow\) in the \(\bar{\alpha}\)-th normal reservoir. \(\mu_{\rm nm}\) and \(t_{\rm nm}\) are the chemical potential and the nearest-neighbor hopping amplitude in normal reservoir, respectively. \(N_{\rm nm}\) represents the total number of sites in the normal reservoirs.
	
	The tight binding Hamiltonian for the 2-node WSM slab in the BDG basis can be written as follows:
	\begin{widetext}
		\begin{eqnarray}\nonumber\label{eq:H2nFT}
			\mathbb{H}_{2n} &=& \frac{1}{2}\sum_{k_W}\sum_{i,i'=1}^{N_{\parallel}}\sum_{jj'=1}^{N_{\perp}}\sum_{\tau\tau'\s\s'}   \Psi^{\dagger}_{ij\tau\s}\tau^{0}_{\tau\tau'}\sigma^{x}_{\s\s'}\Big[\big(2t_x\big(\cos k_W-\cos k_0\big)+2m_{2n}\big)\delta_{i,i'}\delta_{j,j'}-\Big(\frac{m_{2n}}{2}\tau^{0}_{\tau\tau'}\sigma^{x}_{\s\s'}\pm i\tau^{z}_{\tau\tau'}\sigma^{y}_{\s\s'}t\Big)\\
			&&\delta_{i',i\pm1}\delta_{j,j'}-\Big(\frac{m_{2n}}{2}\tau_{0}^{\tau\tau'}\sigma^{x}_{\s\s'}\pm i\tau^{z}_{\tau\tau'}\sigma^{z}_{\s\s'}t\Big)\delta_{i',i}\delta_{j',j\pm 1}\Big]\Psi_{i'j'\tau'\s'}.
			\\\nonumber\label{eq:H4nFT}
			\mathbb{H}_{4n} &=& \frac{1}{2}\sum_{k_W}\sum_{i,i'=1}^{N_{\parallel}}\sum_{jj'=1}^{N_{\perp}}\sum_{\s\s\tau\tau''ss'} \Psi^{\dagger}_{i,j,s,\s}\Big[\big(\lambda s^
			0_{ss'} \tau^0_{\tau\tau'}\sigma^y_{\s\s'} \sin k_W+\big(m_{4n}+2\big)s^y_{ss'}\tau^0_{\tau\tau'}\sigma^y_{\s\s'}\big)\delta_{i,i'}\delta_{j,j'}-\frac{1}{2}\
			\Big(s^y_{ss'}\tau^0_{\tau\tau'}s^y_{\s\s'}\\
			&&\pm i\lambda  s^0_{ss'}\tau^z_{\tau\tau'}\sigma^x_{\s\s'}\Big)\delta_{i',i\pm1}\delta_{j,j'}-\frac{1}{2}\Big(s^y_{ss'}\tau^0_{\tau\tau'}\sigma^y_{\s\s'}\pm i\lambda s^0_{ss'}\tau^z_{\tau\tau'}\sigma^x_{\s\s'}\Big)\delta_{i',i\pm1}\delta_{j,j'}\Big]\Psi_{i',j',s'\s'}.
		\end{eqnarray}
	\end{widetext}
	\twocolumngrid
	Here, the Nambu spinors are defined as \(\Psi_{ij}^{\dagger} = (\psi_{\uparrow}^{\dagger}, \psi_{\downarrow}^{\dagger}, \psi_{\downarrow}, -\psi_{\uparrow}^{\dagger})_{ij}\) and \(\Psi_{ij}^{\dagger} = (\psi_{1\uparrow}^{\dagger}, \psi_{1\downarrow}^{\dagger}, \psi_{2\uparrow}^{\dagger}, \psi_{2\downarrow}^{\dagger}, \psi_{1\downarrow}, -\psi_{1\uparrow}^{\dagger}, \psi_{2\downarrow}, -\psi_{2\uparrow}^{\dagger})_{ij}\), in equations (\ref{eq:H2nFT}) and (\ref{eq:H4nFT}), respectively. \(i\) and \(j\) correspond to the site indices of the slab geometry, where \(i\) is along the \(x_{\parallel}\) direction, and \(j\) is along the \(x_{\perp}\) direction.  \(N_{\parallel}\) and \(N_{\perp}\) represent the number of sites in the \(x_{\parallel}\) and \(x_{\perp}\) directions, respectively.
	
	To establish a connection between the reservoirs and the surfaces of the Weyl semimetal slab, we introduce tunneling matrix Hamiltonians as described by:
	\begin{align}\label{eq:HTb}
		\mathbb{H}_T^{\beta,\alpha'} = \frac{1}{2} \Psi^\dagger_{r} V^{\beta,\alpha'}_{rr'} \Phi^{\alpha'}_{r'} + \text{h.c.},
	\end{align}
	The matrix elements \(V^{\beta,\alpha'}_{rr'}\) contain details about the tunneling links specific to sites identified by \((r,r')\). \(\beta=2n\) and \(4n\) specify the coupling in the case of the 2-node and 4-node WSM, respectively. \(r\) and \(r'\) correspond to coordinates within the WSM and the \(\alpha'\)-th reservoir, respectively.
	
	For the Josephson junction positioned along the top surface, the values of \(\{r, r'\}\) are set to \(\{(L_\perp, 1),N_{\rm sc}\}\) and \(\{(L_\perp, L_{\parallel}),1\}\), indicating the connections with the left and right superconducting reservoirs, respectively. The phase of these left and right superconducting reservoirs are given by \(\phi^L\) and \(\phi^R\), respectively. The superconducting phase difference is given as \(\delta\phi=\phi^R-\phi^L\). For the grounded case, the values of \(\{r, r'\}\) are set to \(\{(1,1),N_{\rm nm}\}\) and \(\{(1, L_{\parallel}),1\}\), indicating the connections with the left and right normal reservoirs, respectively. 
	
	In the cases of the Josephson junction of the 2-node and 4-node WSM, the tunneling matrix elements are given by:
	\begin{eqnarray}
		\label{eq:V2n}
		V^{2n,\alpha'}_{r,r'} &=& t_{\alpha'} \delta_{r,r'}\tau_z\s_0
		\\\label{eq:V4n}
		V^{4n,\alpha'}_{r,r'} &=& t_{\alpha'} \delta_{r,r'} 
		\begin{pmatrix}
			\tau_z\s_0 & 0 \\
			0 & \tau_z\s_0 \\
		\end{pmatrix}
	\end{eqnarray}
	The Green's function of the full system (WSM slab connected with reservoirs), is defined as follows:
	\begin{align}
		G_{\beta}(\omega) = \left(\omega \mathbb{I} - \mathbb{H}_{\beta} - \sum_{\alpha'}\Sigma^{\beta}_{\alpha'}(\omega) \right)^{-1}.
	\end{align}
	Here, \(\mathbb{H}_{\beta}\) is given by Eq.~(\ref{eq:H2nFT}) and Eq.~(\ref{eq:H4nFT}) in the case of the 2-node and 4-node WSM, respectively. \(\Sigma^{\beta}_{\alpha'}(\omega)= V^{\beta,\alpha'}g^{\alpha'}(\omega)V^{{\beta,\alpha'}\dagger}\) is the self-energy term corresponding to the \(\alpha'\) reservoir, where \(g^{\alpha'}(\omega)\) represents the Green's functions for the \(\alpha'\)-th reservoir.
	
	In this system, the bond current flow at site \((i,j)\) can be expressed as \(\hat{J}^{\beta}_{ij}=(ie/\hbar)\left[\mathbb{H}_{\beta},N^{\beta}_{ij}\right]\), where \(N^{\beta}_{ij}\) represents the electronic number operator at site \((i,j)\). For the 2-node WSM, the net current flow along the \(x_{\parallel}\) and \(x_{\perp}\) directions is given by the following expressions:
	
	\begin{eqnarray}\label{eq:j2n_||}\nonumber
		\la \hat{J}^{2n}_{(i\rightarrow i+1,j)}\ra&=&-\frac{e}{\hbar}\sum_{k_W\s\s'}\Big(\frac{m_{2n}}{2}\text{Im}
		\left[\la \psi^{\dagger}_{i,j,\s}\psi_{i+1,j,\s'}\ra\right]+\,t\sigma'\\
		&&\text{Re}\left[\la \psi^{\dagger}_{i,j,\s}\psi_{i+1,j,\s'}\ra\right]\Big),
	\end{eqnarray}	
	\begin{equation}
		\langle \hat{J}^{2n}_{(i,j\rightarrow j+1)} \rangle =-\frac{e}{\hbar} \sum_{k_W\s\s'} \left(\frac{m_{2n}}{2} - t\sigma\right)\text{Im}\left[ \langle \psi^{\dagger}_{i,j,\s}\psi_{i,j+1,\s'}\rangle\right].\label{eq:j2n_perp}
	\end{equation}
	
	Similarly, for the 4-node WSM, the net current flow along the \(x_{\parallel}\) and \(x_{\perp}\) directions is given by the following expressions:
	
	\begin{eqnarray}\label{eq:j4n_||}\nonumber
		\centering
		\la \hat{J}^{4n}_{(i\rightarrow i+1,j)}\ra&=&-\frac{e}{2\hbar}\sum_{k_Ws\s\s'}\Big(\lambda\s'\text{Re}
		\left[\la \psi^{\dagger}_{i,j,s,\s}\psi_{i+1,j,s,\s'}\ra\right]\\
		&&+\,s\s'\text{Im}
		\left[\la \psi^{\dagger}_{i,j,s,\s}\psi_{i+1,j,s',\s'}\ra\right]\Big),\\\nonumber
		%
		\nonumber
		\langle \hat{J}^{4n}_{(i,j\rightarrow j+1)} \rangle& =& -\frac{e}{2\hbar} \sum_{k_Ws\s\s'} \Big( \lambda \text{Re} \left[\langle \psi^{\dagger}_{i,j,s,\s}\psi_{i,j+1,s,\s'} \rangle\right]\\	\label{eq:j4n_perp}
		&& +\,s\s' \text{Im}\left[\langle \psi^{\dagger}_{i,j,s,\s}\psi_{i,j+1,s',\s'} \rangle\right]\Big).
	\end{eqnarray}
	In the above mentioned equations, \(\sigma'=+1\) and \(-1\) represent the spin-up and spin-down cases, respectively. Similarly, \(s=+1\) and \(-1\) correspond to orbital-1 and orbital-2 cases, respectively. The symbol \(\langle . \rangle\) signifies the thermal average taken over the states of the reservoir. The averages \(\langle . \rangle\) in the above equations can be computed using the Non-Equilibrium Green's Function approach~\cite{martin1994microscopic} as:
	
	\begin{equation}
		G^{+-}_{\beta,r'r}(t,t') = i \begin{pmatrix}
			\langle \psi^{\dagger}_{r}(t)\psi_{r'}(t') \rangle & \langle \psi_{r}(t)\psi_{r'}(t') \rangle \\
			\langle \psi^{\dagger}_{r}(t)\psi^{\dagger}_{r'}(t') \rangle & \langle \psi_{r}(t)\psi^{\dagger}_{r'}(t') \rangle
		\end{pmatrix}
	\end{equation}
	In this equation, \(r\) and \(r'\) represent site indices encompassing the information of the corresponding local degree of freedom. Using Fourier transform, we can write:
	\begin{align}
		G_{\beta}^{+-}(\omega)=f(\omega)\left[G_{\beta}^A(\omega)-G_{\beta}^R(\omega)\right].
	\end{align}    
	In the above equation, \(G_{\beta}^A(\omega)\) and \(G_{\beta}^R(\omega)\) refer to the advanced and retarded Green's functions, respectively. 
	
	The bond currents along distinct bonds in the \(x_{\parallel}\) and \(x_{\perp}\) directions are calculated using Eq.~(\ref{eq:j2n_||}) and Eq.~(\ref{eq:j2n_perp}) in the 2-node case, and Eq.~(\ref{eq:j4n_||}) and Eq.~(\ref{eq:j4n_perp}) in the 4-node case, respectively. The net Josephson current along the \(x_{\parallel}\) direction is obtained by summing over the index-\(j\) from $1$ to $N_{\perp}$, in Eq.~(\ref{eq:j2n_||}) and Eq.~(\ref{eq:j4n_||}) for the 2-node and 4-node cases, respectively.
	
	\section{Symmetry Analysis}\label{sec:Symmetry-Analysis}
	This section explains the presence of anomalous current using the symmetry analysis of the individual of full Hamiltonian (WSM slab connected with reservoirs) in the 2-node and 4-node cases.	In the case of 2-node WSM slab, these terms can be expressed as: 
	\begin{align}\nonumber
		\centering
		h^{2n,M} & = \big(2t_x\big(\cos k_W - \cos k_0\big) + 2m_{2n}\big)\mathbf I_N\otimes \tau_0\otimes\s_x\\
		&\hspace{4mm} - \frac{m_{2n}}{2}\Big(\mathcal{A} +\mathcal{B}\Big)\otimes \tau_{0}\otimes\s_{x}
		\\\label{eq:H2n_2}
		h^{2n,||} & = \pm it\mathcal{A}\otimes\tau_{z}\otimes \s_{x}.\\
		h^{2n,\perp} & =\pm it\mathcal{B}\otimes\tau_{z}\otimes \s_{x}.
	\end{align}
	Here, $h^{2n,M}$ represents the diagonal and off-diagonal contributions of the mass term, whereas $h^{2n,||}$ and $h^{2n,\perp}$ correspond to the off-diagonal spin-orbit terms along the $i$ ($x_{\parallel}$) and $j$ ($x_{\perp}$) directions, respectively. Additionally, the matrices $\mathcal{A}_{(i,j),(i',j')}=\delta_{i',i\mp1}\delta_{j,j'}$, $\mathcal{B}_{(i,j),(i',j')}=\delta_{j',j\mp1}\delta_{i,i'}$ and $\mathbf I_{N}$ is the identity matrix of size $N$. For the WSM slab $N$ is given as $(N_{\parallel}\times N_{\perp})$. For the superconductors:
	\begin{align}\label{eq:H2n_3}\nonumber
		h_{{\rm sc}}(\delta\phi) & = \sum_{\alpha}\Big[ \mu_{\rm sc} I_N\otimes \tau_z\otimes\s_0 + \Delta_{\rm sc} I_N\otimes (\tau_x \cos\phi^{\alpha} - \\
		&\hspace{4mm} \tau_y \sin\phi^{\alpha})\otimes\s_0 + t_{\rm sc} \mathcal{A}\otimes\tau_z\otimes\s_0 \Big]
	\end{align}
	In addition to this terms, \(\sum_{\alpha,\beta}V^{\beta,\alpha}_{rr'}\) correspond to the tunneling terms which connects the reservoirs at the surfaces of WSM slab as defined in Eq.~(\ref{eq:V2n}). Furthermore, the Zeeman field terms \(\tilde{b}^{2n}_W\), \(\tilde{b}^{2n}_{\parallel}\), and \(\tilde{b}^{2n}_{\perp}\) in the BDG basis, are represented by \(b_W\tau_0\s_x\), \(b_{\parallel}\tau_0\s_y\), and \(b_{\perp}\tau_0\s_z\).
	Similarly, in the 4-node case,
	\begin{align}
		\centering
		h^{4n,d} &=  i\lambda\sin k_y \mathbf I_N\otimes s_0\otimes \tau_0\otimes\s_y;\\\label{eq:H4n_2}\nonumber
		h^{4n,M} &= (m_{4n}+2)\mathbf I_N\otimes s_y\otimes \tau_0\otimes\s_y - \frac12\Big(\mathcal{A} +\mathcal{B}\Big)\otimes
		\\
		&\hspace{4mm} s_y\otimes \tau_0\otimes\s_y.\\\label{eq:H4n_3}
		h^{4n,||} &= \pm i\lambda\mathcal{A}\otimes s_0\otimes\tau_{z}\otimes \s_{x}.\\\label{eq:H4n_4}
		h^{4n,\perp} &=\pm i\lambda\mathcal{B}\otimes s_0\otimes\tau_{z}\otimes \s_{x}.
	\end{align}
	Here, $h^{4n,d}$ and $h^{4n,M}$ represents the diagonal and off-diagonal contributions of the mass term, whereas $h^{4n,||}$ and $h^{4n,\perp}$ correspond to the off-diagonal spin-orbit terms along the $i$ ($x_{\parallel}$) and $j$ ($x_{\perp}$) directions, respectively. Additionally, the matrices. In addition to this terms, \(\sum_{\alpha,\beta}V^{\beta,\alpha}_{rr'}\) correspond to the tunneling terms which connects the reservoirs at the surfaces of WSM slab as defined in Eq.~(\ref{eq:V4n}). Furthermore, the Zeeman field terms \(\tilde{b}^{4n}_W\), \(\tilde{b}^{4n}_{\parallel}\), and \(\tilde{b}^{4n}_{\perp}\) in the BDG basis, are represented by $b_Ws_0\otimes\tau_0\otimes\s_y$, $b_{\parallel}s_0\otimes\tau_0\otimes\s_x$, and $ b_{\perp}s_0\otimes\tau_0\otimes\s_z$.
	
	The symmetries of these individual terms in the in the case of 2-node WSM can be expressed as: 
	\begin{align}\label{eq:slist2n}
		\begin{bmatrix}
			. &h^{2n,M} & h^{2n,||} & h^{2n,\perp} & \tilde{b}^{2n}_W & \tilde{b}^{2n}_{\parallel} & \tilde{b}^{2n}_{\perp} & h_{{\rm sc}}(\delta\phi) \\
			\rowcolor[gray]{.8} \mathcal{\tilde{T}}_{2n} & - & + & + & - & - & - & \delta\phi \rightarrow -\delta\phi \\
			\tilde{\s}_x & + & - & - & + & - & - & + \\
			\rowcolor[gray]{.8}\tilde{\s}_y & - & + & - & - & + & - & + \\
			\tilde{\s}_z & - & - & + & - & - & + & + \\
			\tilde{\mathcal{R}}_{\parallel} & + & - & + & + & + & + & \delta\phi \rightarrow -\delta\phi \\
			\rowcolor[gray]{.8}\tilde{\mathcal{R}}_{\perp} & + & + & - & + & + & + & +
		\end{bmatrix}
	\end{align}
	Here, the symbols `$+$' and `$-$' represent the signs accumulated in given terms in first row under the symmetry operators given in the corresponding entries in the first column. These symmetries include time-reversal indicated as $\tilde{\mathcal{T}}_{2n}=\mathbf{I}_N\otimes\mathcal{T}_{2n}$. Spatial inversions along $||$ and $\perp$-directions, denoted as $\tilde{\mathcal{R}}_{\parallel}=\mathbf{I}_N\otimes\tau_0\otimes\s_0\otimes\mathcal{R}_{\parallel}$ and $\tilde{\mathcal{R}}_{\perp}=\mathbf{I}_N\otimes\tau_0\otimes\s_0\otimes\mathcal{R}_{\perp}$. Spin rotations denoted as $\tilde{\s}_x=\mathbf{I}_N\otimes\tau_0\otimes\sigma_x$, $\tilde{\s}_y=\mathbf{I}_N\otimes\tau_0\otimes\sigma_y$, and $\tilde{\s}_z=\mathbf{I}_N\otimes\tau_0\otimes\sigma_z$.
	
	The symmetries of these individual terms in the in the case of 4-node WSM can be expressed as: 
	\begin{align}\label{eq:slist4n}
		\begin{bmatrix}
			\text{.} & h^{4n,M} & h^{4n,d} & h^{2n,||} &h^{2n,\perp}  & \tilde{b}^{4n}_{\parallel} & \tilde{b}^{4n}_{W} & \tilde{b}^{4n}_{\perp} & h_{\text{sc}}(\delta    \phi) \\
			\rowcolor[gray]{.8} \tilde{\mathcal{T}}_{4n} & + & + & + & + & - & - & - & \delta\phi \rightarrow -\delta\phi \\
			\rowcolor[gray]{.8}\tilde{\s}_x & - & - & + & - & + & - & - & + \\
			\tilde{\s}_y & + & + & - & - & - & + & - & + \\
			\tilde{\s}_z & - & - & - & + & - & - & + & + \\
			\tilde{s}_x & - & + & + & + & + & + & + & + \\
			\tilde{s}_y & + & + & + & + & + & + & + & + \\
			\rowcolor[gray]{.8}\tilde{s}_z & - & + & + & + & + & + & + & + \\
			\mathcal{R}_{\parallel} & + & + & - & + & + & + & + & \delta\phi \rightarrow -\delta\phi \\
			\rowcolor[gray]{.8}\mathcal{R}_{\perp} & + & + & + & - & + & + & + & +
		\end{bmatrix}
	\end{align}
	Here, these symmetries include time-reversal indicated as $\tilde{\mathcal{T}}_{4n}=\mathbf{I}_N\otimes\mathcal{T}_{4n}$. Spatial inversions along $||$ and $\perp$-directions, denoted as $\tilde{\mathcal{R}}_{\parallel}=\mathbf{I}_N\otimes s_0\otimes\tau_0\otimes\s_0\otimes\mathcal{R}_{\parallel}$ and $\tilde{\mathcal{R}}_{\perp}=\mathbf{I}_N\otimes s_0\otimes\tau_0\otimes\s_0\otimes\mathcal{R}_{\perp}$. Spin rotations denoted as $\tilde{\s}_x=\mathbf{I}_N\otimes s_0\otimes\tau_0\otimes\sigma_x$, $\tilde{\s}_y=\mathbf{I}_N\otimes s_0\otimes\tau_0\otimes\sigma_y$, and $\tilde{\s}_z=\mathbf{I}_N\otimes s_0\otimes\tau_0\otimes\sigma_z$.
	
	Using Eq.~(\ref{eq:slist2n}) and Eq.~(\ref{eq:slist4n}), for 2-node and 4-node cases we have defined the symmetry operators \(\mathcal{P}_{2n} = \mathcal{T}_{2n} \sigma_y \mathcal{R}_{\perp}\) and \(\mathcal{P}_{4n} = \mathcal{T}_{4n}|_{{k_W}\rightarrow{-k_W}} s_z \sigma_x \mathcal{R}_{\perp}\), respectively. This symmetry gives rise to \(\mathcal{P}_{\beta} \mathbb{H}^{\beta}_{\text{full}}(\Phi^L,\Phi^R) \mathcal{P}_{\beta}^{-1} = \mathbb{H}^{\beta}_{\text{full}}(-\Phi^L,-\Phi^R)\), where \(\mathbb{H}^{\beta}_{\text{full}}\) represents the full Hamiltonian of the Josephson junction. This implies \(E(\phi)=E(-\phi)\) and ensures \(I(\phi)=-I(-\phi)\), ensuring the absence of anomalous current.
	
	When the Zeeman field \(b_{\parallel}\) is applied, we observe that this Zeeman field term disrupts this symmetry, resulting in the emergence of anomalous current in both WSM cases. This symmetry breaking leads to \(E_{n}(\phi^L,\phi^R) \neq E_{n}(-\phi^L,-\phi^R)\), resulting in \(I(\phi) \neq -I(-\phi)\) and causing the presence of anomalous current. To understand this anomalous current and oscillatory behavior further, we employed a network model study given in following section.
	
\end{document}